\documentclass[fleqn,twocolumns,usenatbib]{mnras}

\usepackage[T1]{fontenc}
\usepackage{ae,aecompl}

\usepackage{graphicx}	
\usepackage{amsmath}	

\usepackage{amssymb}	

\title[Si ISM X-ray absorption]{Silicon ISM X-ray absorption: the gaseous component}

\author[Gatuzz et al.]{
E. Gatuzz$^{1}$\thanks{E-mail: egatuzz@mpe.mpg.de},
T. W. Gorczyca$^{2}$,
M. F. Hasoglu$^{3}$,
N. S. Schulz$^{4}$,\newauthor
L. Corrales$^{5}$,
and C. Mendoza$^{2}$
\\
$^{1}$Max-Planck-Institut f\"ur extraterrestrische Physik, Gie{\ss}enbachstra{\ss}e 1, 85748 Garching, Germany\\
$^{2}$Department of Physics, Western Michigan University, Kalamazoo, MI 49008, USA\\
$^{3}$Department of Computer Engineering, Hasan Kalyoncu University, 27100 Sahinbey, Gaziantep, Turkey\\
$^{4}$Kavli Institute for Astrophysics and Space Research, Massachusetts Institute of Technology, Cambridge, MA 02139 \\
$^{5}$Department of Astronomy, University of Michigan, Ann Arbor, MI 48109, USA\\
}

\date{Accepted XXX. Received YYY; in original form ZZZ}
\pubyear{2020}
\begin{document}
 \label{firstpage}
\pagerange{\pageref{firstpage}--\pageref{lastpage}}
\maketitle
\begin{abstract}
We present a detailed analysis of the gaseous component of the Si K~edge using high-resolution {\it Chandra} spectra of low-mass X-ray binaries. We fit the spectra with a modified version of the {\it ISMabs} model, including new photoabsorption cross sections computed for all Si ionic species. We estimate column densities for {\rm Si}~{\sc i}, {\rm Si}~{\sc ii}, {\rm Si}~{\sc iii}, {\rm Si}~{\sc xii} and {\rm Si}~{\sc xiii}, which trace the warm, intermediate temperature and hot phases of the Galactic interstellar medium. We find that the ionic fractions of the first two phases are similar. This may be due to the physical state of the plasma determined by the temperature or to the presence of absorber material in the close vicinity of the sources. Our findings highlight the need for accurate modeling of the gaseous component before attempting to address the solid component.
\end{abstract}


\section{Introduction}\label{sec_in}

The interstellar medium (ISM) is one of the most important components in stellar evolution because it regulates Galactic star formation and cooling rates \citep{won02,big08,ler08,lad10,lil13}. The ISM displays a multiphase structure that is temperature dependent \citep{wad01,ton09,rup13,zhu16}. High-resolution X-ray spectra enable the study of such environments through the analysis of the absorption features from the gas in the line-of-sight between the source and observer \citep{jue04,yao09,lia13,luo14,sch16,gat16,joa16,gat18a}.

Silicon can be used to probe the phase transitions between gas and solid states. This element can be found in interstellar clouds \citep{hen09}, the circumstellar environment of oxygen-rich asymptotic giant branch (AGB) stars \citep{gai09}, protostellar disks \citep{apa10} and comets \citep{han10}. \citet{sch16} analyzed the Si K-edge absorption region with a phenomenological model using high-resolution {\it Chandra} spectra of 11 low-mass X-ray binaries (LMXBs), and compared the results with dust model calculations for olivine by \citet{dra03b}. They found multiple photoabsorption edge features including a variable warm absorber from ionized atomic silicon (e.g. {\rm Si}~{\sc xiii}), which they believe is associated to material intrinsic to the sources. \citet{zee19} analyzed the interstellar dust absorption and scattering in the Si K~edge using laboratory measurements of several silicate compounds. They found that most of the interstellar dust can be modeled by amorphous olivine, concluding there is no relation between depletion and extinction along the line of sight.

We present an analysis of the Si K-edge absorption region using {\it Chandra} observation of 16 LMXBs. We specifically focus on the atomic component including the neutral as well as ionized species, exploring their presence and fractions in the Si K~edge without excluding co-existing dust components. In Section~\ref{sec_xray_data} we describe the data sample and the spectral fitting procedure. We discuss the results obtained from the fits in Section~\ref{sec_dis}. Finally, we summarize the main results of our analysis in Section~\ref{sec_con}.

\section{X-ray observations and spectral fitting}\label{sec_xray_data}

To compile the data sample we have selected {\it Chandra} spectra of LMXBs with a high number of counts in the Si K-edge wavelength region (i.e. more than 1000 counts in the range 6--7~\AA). We did not impose constraints in the significant detection of a particular line (e.g. {\rm Si}~{\sc xiii}) in order to get an unbiased sample. A total of 16 Galactic sources were selected. Table~\ref{tab_data} lists the {\it Chandra} observations used in the analysis including their Galactic coordinates, distances and ${\rm HI}$ column densities obtained from \citet{wil13}. Although such column densities do not fully reflect the X-ray column density values given the small wavelength region analyzed in our sample ($\sim 1$ \AA), they can be safely used.

\begin{table}
\scriptsize
\caption{\label{tab_data}List of {\it Chandra} HETG observations.}
\centering
\begin{tabular}{cccccc}
\hline
Source    & Galactic    &Distance& $N({\rm HI})$ \\
& Coordinates &(kpc)   & ($10^{21}$ cm$^{-2}$) \\
\hline
4U~0614+091  &$(200.87,-3.36)$&$2.2\pm 0.7^{a}$& $5.86$  \\
4U~1626-67   &$(321.78,-13.09)$&$3.5^{+0.2}_{-0.3}$ $^{b}$& $1.30$  \\
4U~1636-53   &$(332.91,-4.81)$& $6\pm 0.5^{d}$& $4.04$  \\
4U~1705-44   &$(343.32,-2.34)$&$7.6\pm 0.3^{d}$& $8.37$  \\
4U~1728-34 	 &$(354.30,-0.15)$&$5.2\pm 0.5^{d}$&$13.9$  \\
4U~1735-44   &$(346.05,-6.99)$&$ 9.4\pm 1.4^{e}$&$3.96$  \\
4U~1820-30   &$(2.78,-7.91)$& $7.6\pm 0.4^{f}$&$2.33$ \\
Cygnus~X-2   &$(87.32,-11.31)$&$13.4 \pm 2.0^{e}$&$3.09$  \\
EXO~0748-676 &$(279.97,-19.81)$&$8.0\pm 1.2^{e}$&$1.51$  \\
GX~339-4     &$(338.93,-4.32)$&$10.0\pm 4.5^{g}$&$5.18$ \\
GX~349+2     &$(349.10,2.74)$&$9.2^{c}$&$6.13$  \\
GX~9+9       &$(8.51,9.03)$&$4.4^{c}$&$3.31$  \\
GX~340+00    &$(339.58,-0.07)$&$11.0\pm 0.3^{h}$&$20.8$ \\
GX~5-1       &$(5.08,-1.02)$&$9.2^{h}$&$10.4$ \\
GX~3+1       &$(2.29,0.79)$&$5.0_{-0.7}^{+0.8}$ $^{j}$&$10.7$ \\
GX~13+1      &$(13.51,0.10)$&$7\pm 1^{h}$&$13.6$ \\
\hline
\multicolumn{4}{l}{Distances obtained from $^{a}$\citet{pae01b}; }\\
\multicolumn{4}{l}{$^{b}$\citet{sch19}; $^{c}$\citet{gri02}; }\\
\multicolumn{4}{l}{$^{d}$\citet{gall08}; $^{e}$\citet{jon04};}\\
\multicolumn{4}{l}{$^{f}$\citet{kul03};$^{g}$\citet{hyn04};}\\
\multicolumn{4}{l}{$^{h}$\citet{chr97};$^{j}$\citet{oos01}.}
\end{tabular}
\end{table}

All observations were reduced following the standard Chandra Interactive Analysis of Observations (CIAO, version 4.12) threads\footnote{\url{http://cxc.harvard.edu/ciao/threads/gspec.html}} to obtain the High Energy Grating (HEG) spectra. We used the {\tt findzo} algorithm\footnote{\url{http://space.mit.edu/cxc/analysis/findzo/}} to estimate the zero-order position of the spectra, analyzing a total of 132 observations. For each source, we fitted all observations simultaneously, modeling the continuum with a {\tt powerlaw*constant} model. While the Photon-Index was the same free parameter for all observations, {\tt constant} accounts for differences in the normalization. We note that, given the small wavelength region to analyze, such a model is acceptable even if different observations show different disc accretion states.  The spectral fitting was carried out with the {\sc xspec} package (version 12.10.1\footnote{\url{https://heasarc.gsfc.nasa.gov/xanadu/xspec/}}). Finally, we used {\tt cash} statistics \citep{cas79} without rebinning the spectra.

\begin{figure}
\includegraphics[scale=0.32]{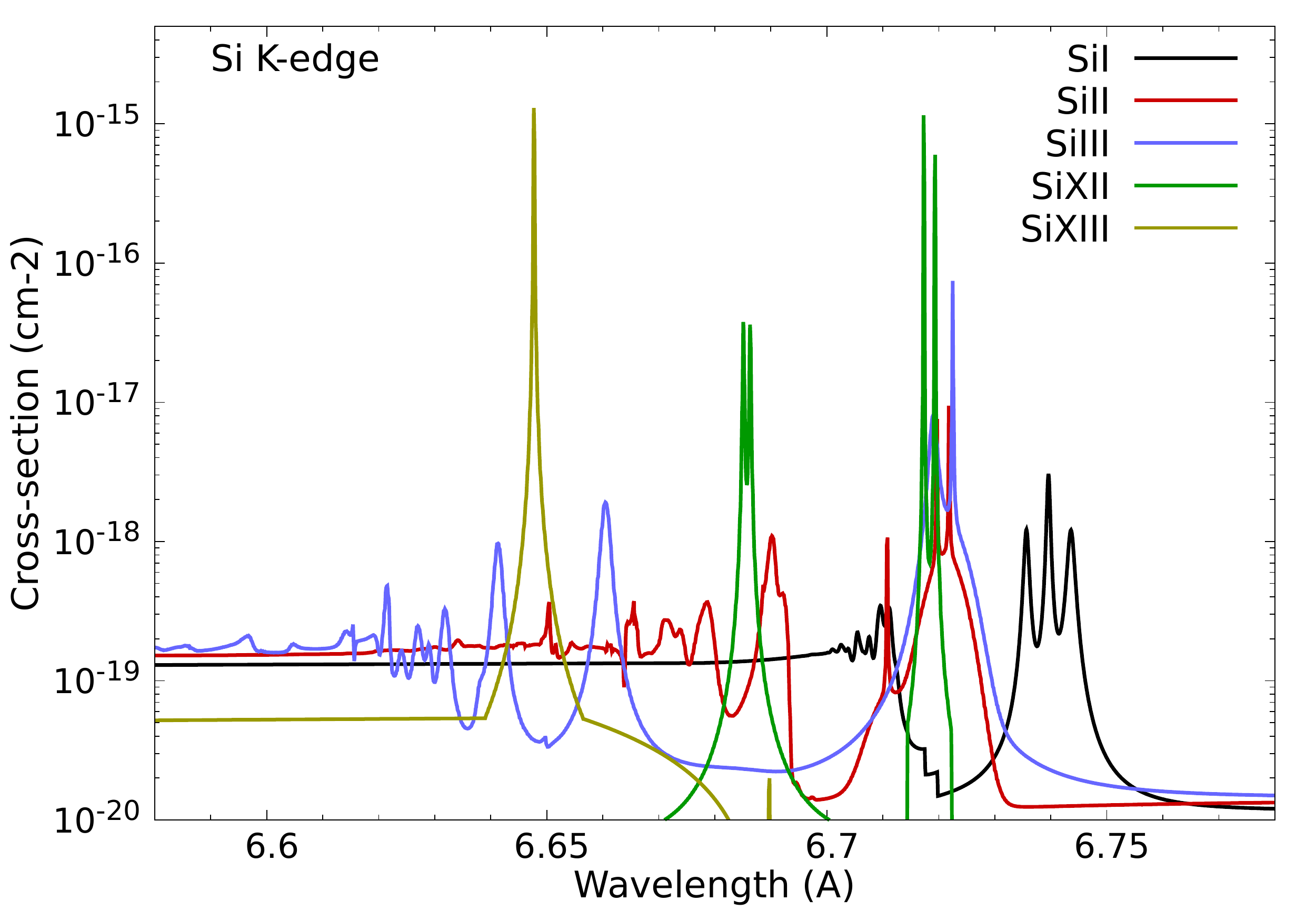}
      \caption{ {\rm Si}~{\sc i}, {\rm Si}~{\sc ii}, {\rm Si}~{\sc iii}, {\rm Si}~{\sc xii} and {\rm Si}~{\sc xiii} photoabsorption cross sections included in the model. }\label{fig_atom}
   \end{figure}

Figure~\ref{fig_atom} shows the {\rm Si}~{\sc i}, {\rm Si}~{\sc ii}, {\rm Si}~{\sc iii}, {\rm Si}~{\sc xii} and {\rm Si}~{\sc xiii} K-edge photoabsorption cross sections used in these models, which correspond to the ions expected in the warm ($10^{4}$~K), intermediate temperature ($10^{4.7}$~K) and hot ($10^{6.3}$~K) phases of the ISM \citep{gat18a}. The {\rm Si}~{\sc i} photoabsorption cross section was computed by Gorczyca et al (2020, in preparation) while those for the ionic species are from \citet{wit09}. We included these Si K-edge photabsorption cross sections in a modified version of the {\it ISMabs} absorption model \citep{gat15}, which allow us to treat the column densities of the Si ions as free parameters.

\begin{figure}
\includegraphics[scale=0.45]{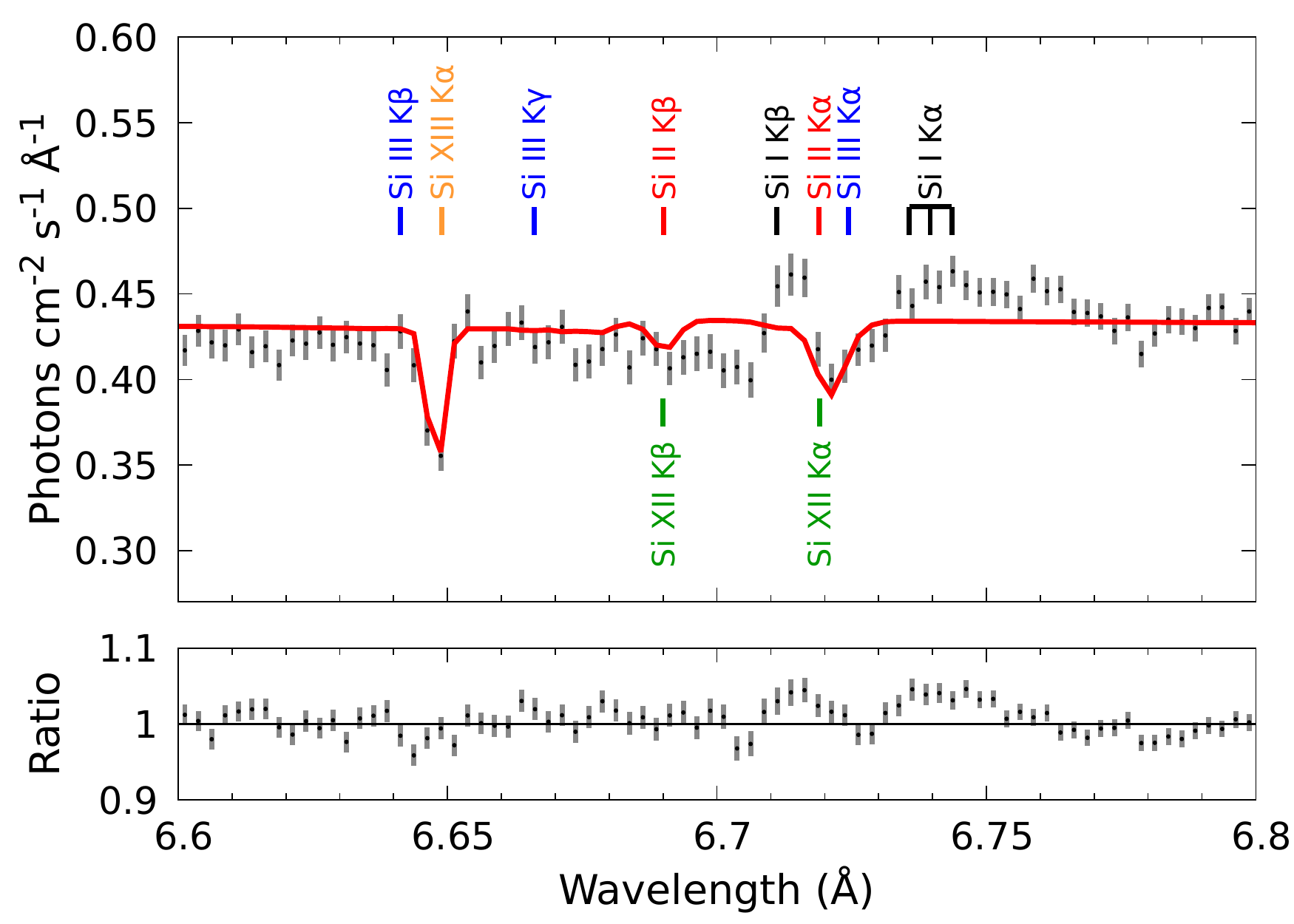}
      \caption{Best fit results after fitting all {\it Chandra} HETG data simultaneously for the {\rm Si} K-edge wavelength region. The main theoretical resonances for the {\rm Si}~{\sc i} (black), {\rm Si}~{\sc ii} (red), {\rm Si}~{\sc iii} (blue), {\rm Si}~{\sc xii} (green) and  {\rm Si}~{\sc xiii} (yellow) are indicated. Residuals in the 6.7--6.75 \AA\ are likely due to dust scattering. }\label{fig_combined}
   \end{figure}

We performed a simultaneous fit of all observations available to estimate mean values for the column densities, as well as to determine the accuracy of the atomic data included in the model. Figure~\ref{fig_combined} shows the HEG spectrum orders $\pm 1$ combined for illustrative purposes. The vertical lines correspond to the main theoretical resonance positions of the Si ions. We have found that the overall wavelength positions in the theoretical cross sections agree with the astronomical observations, although the maximum resolution of the HETG instrument in first order ($\sim 620$) is not high enough to resolve the {\rm Si}~{\sc i} K$\alpha$ triplet.  The best-fit results are listed in Table~\ref{tab_ismabs_combined} showing that the mean column densities for {\rm Si}~{\sc i} and {\rm Si}~{\sc ii} are similar. We also note residuals around the Si~K~edge indicating the contribution of the solid state \citep{sch16}. The model presented here does not include a solid Si dust component, so the {\rm Si}~{\sc i} abundance values obtained in the fits are likely to be overestimated as described in \citet{cor16}. This work also describes how dust scattering near resonant absorption features causes spectral features that mimic emission, such as those seen in the residuals of Figure~\ref{fig_combined}, for high-resolution imaging datasets from {\it Chandra}.

 \begin{table}
\caption{\label{tab_ismabs_combined}Mean ISM silicon column densities obtained for all sources fitted simultaneously. }
\scriptsize
\centering
\begin{tabular}{cccccccc}
\hline
Si\,{\sc i}/H\,{\sc i} &  Si\,{\sc ii}/H\,{\sc i}  &  Si\,{\sc iii}/H\,{\sc i} &  Si\,{\sc xii}/H\,{\sc i} &  Si\,{\sc xiii}/H\,{\sc i}   &{\tt cash}/d.of.\\
\hline
$<1.04$&$0.99\pm 0.13$ &$<0.03$& $<0.03$&$0.23\pm 0.05$& 51573/47073 \\
\hline
\multicolumn{5}{l}{ Column density ratios in units of $10^{-5}$.}
 \end{tabular}
\end{table}

\section{Results and discussion}\label{sec_dis}

\begin{figure*}
        \hspace*{-5mm}
\includegraphics[scale=0.25]{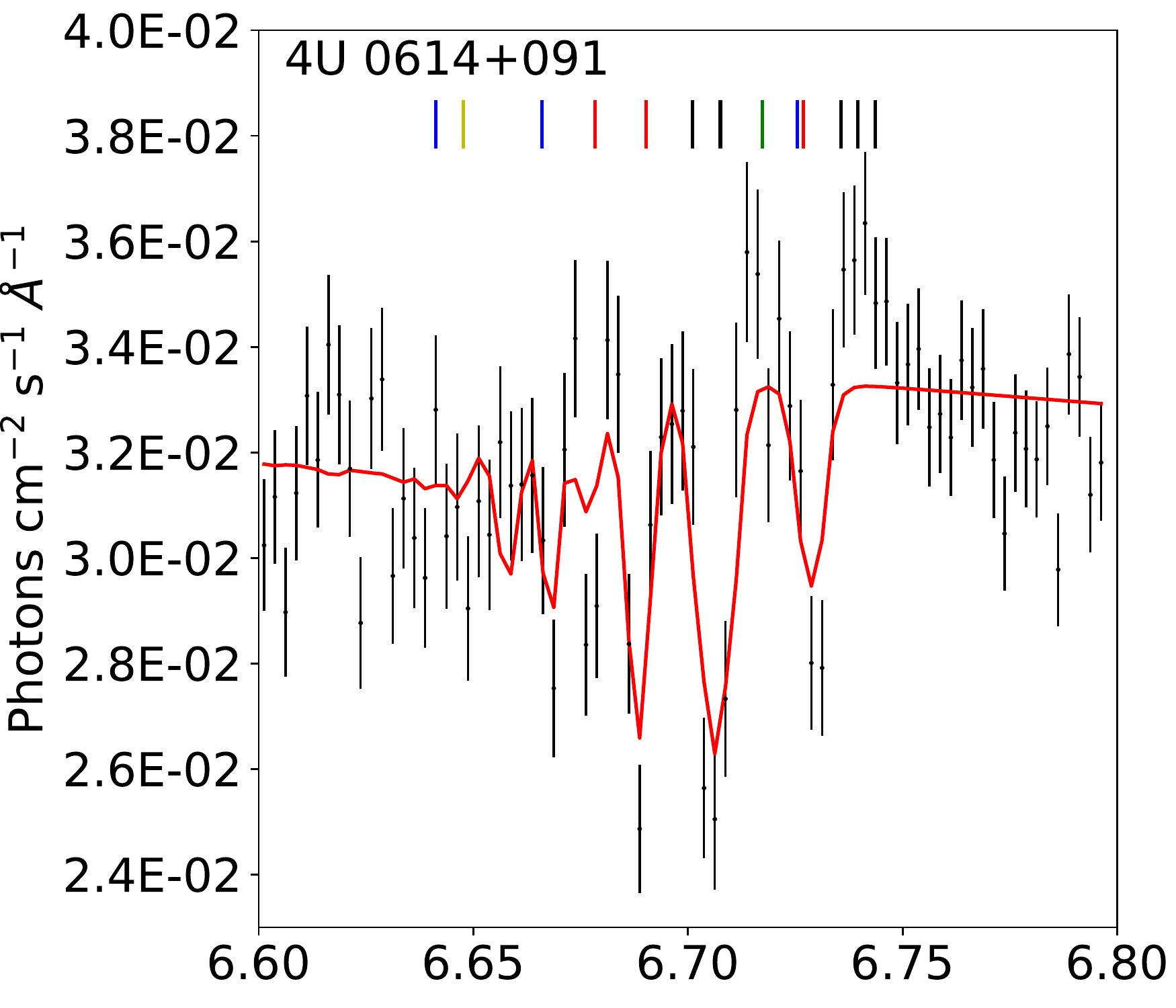}
\includegraphics[scale=0.25]{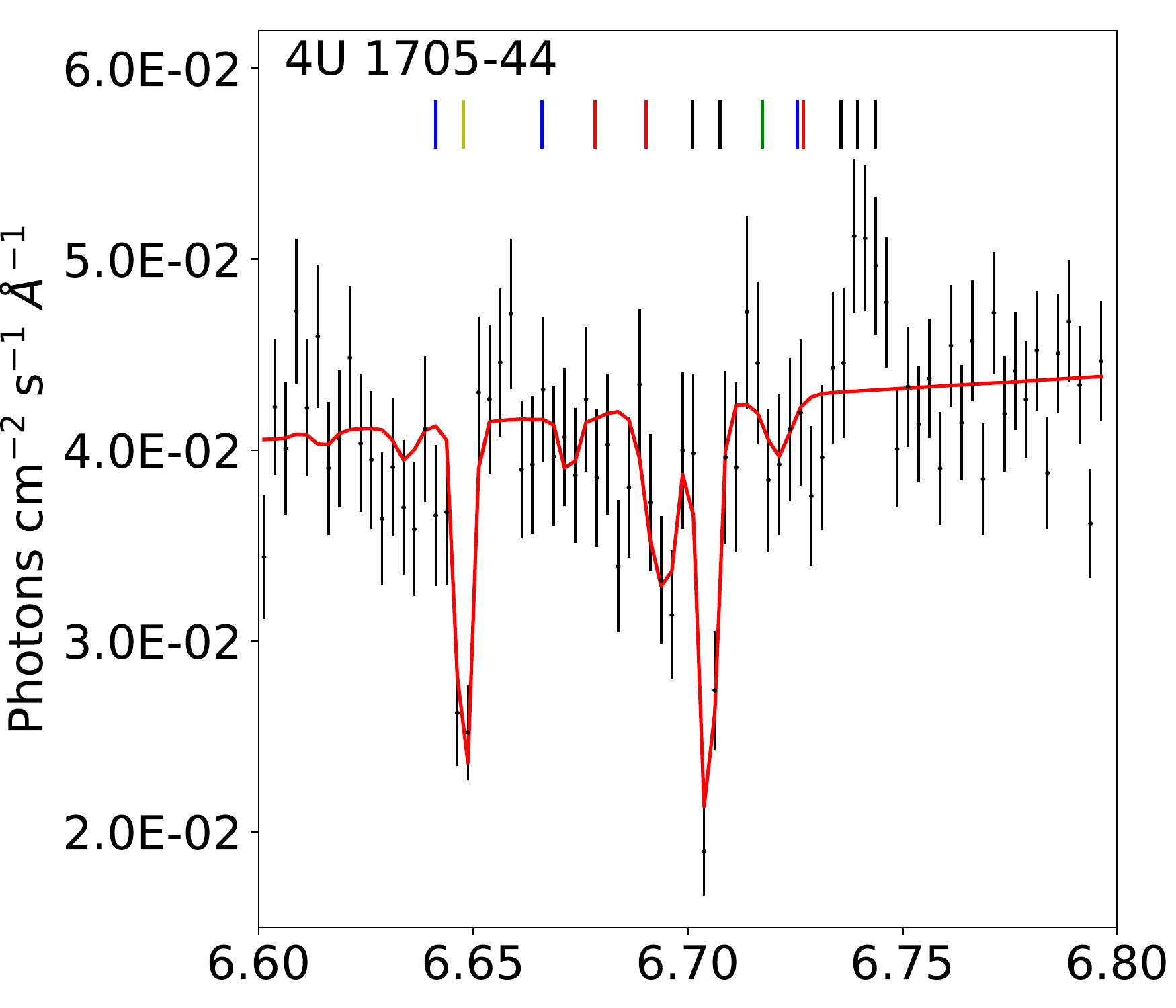}
\includegraphics[scale=0.25]{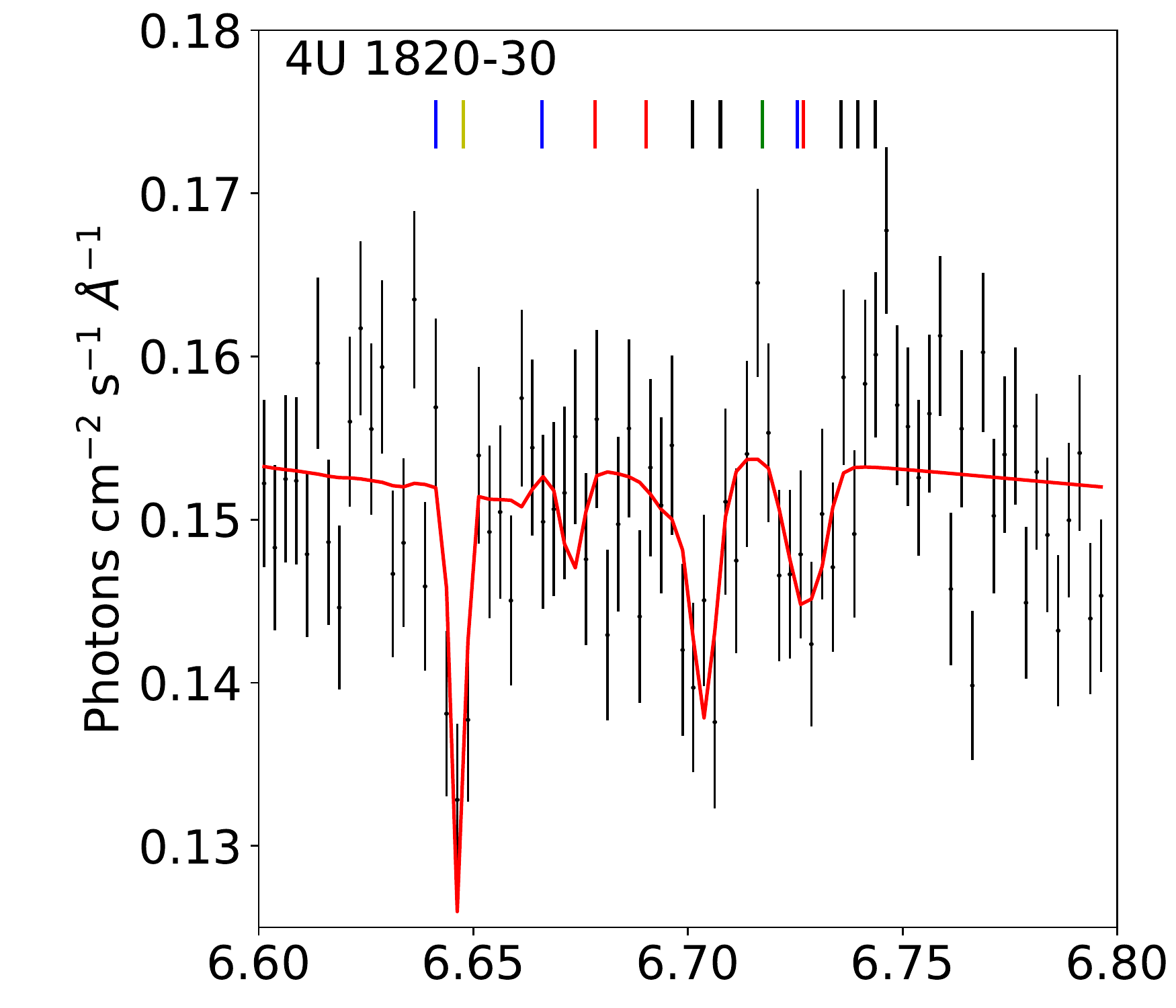}
\includegraphics[scale=0.25]{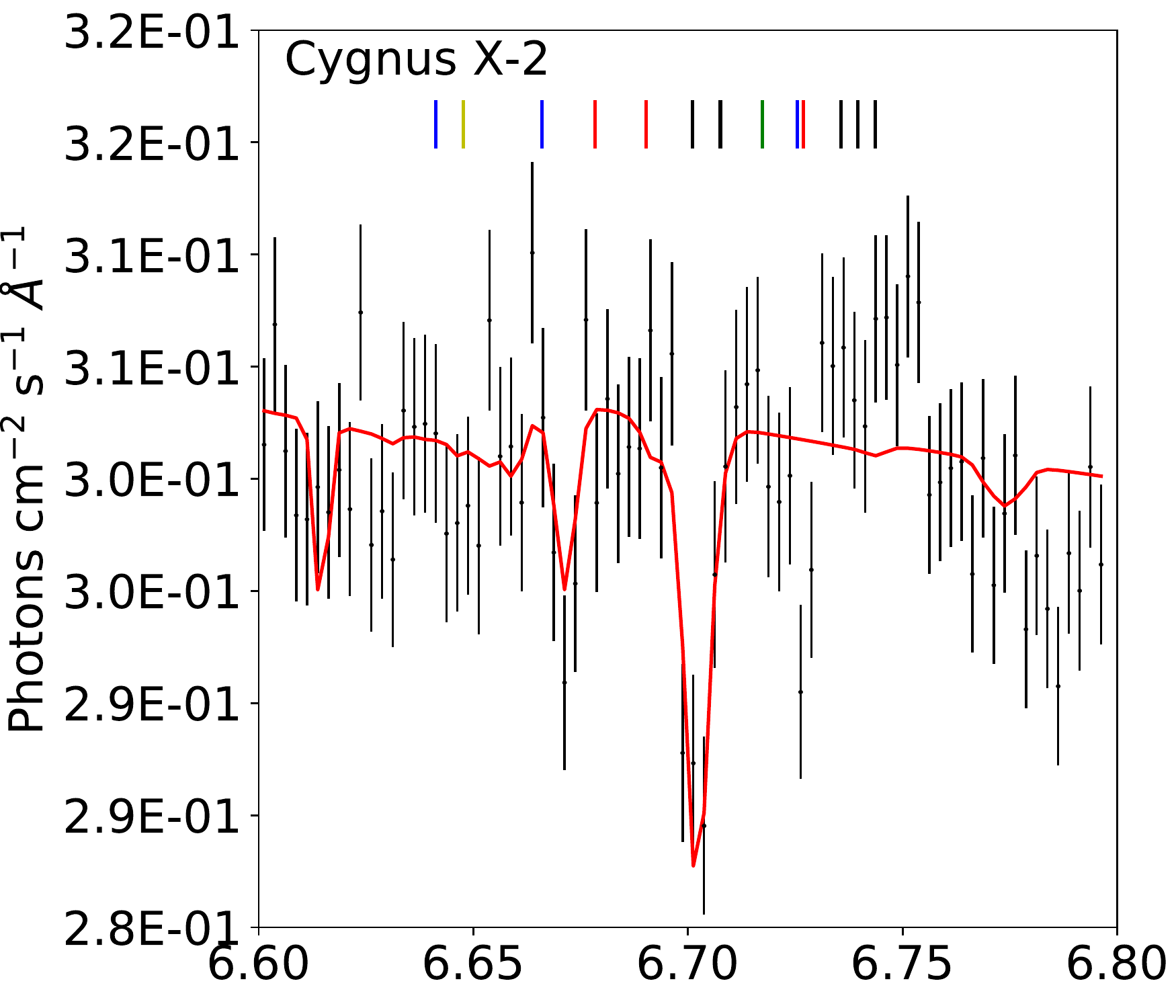}\\
 \hspace*{-5mm}
\includegraphics[scale=0.25]{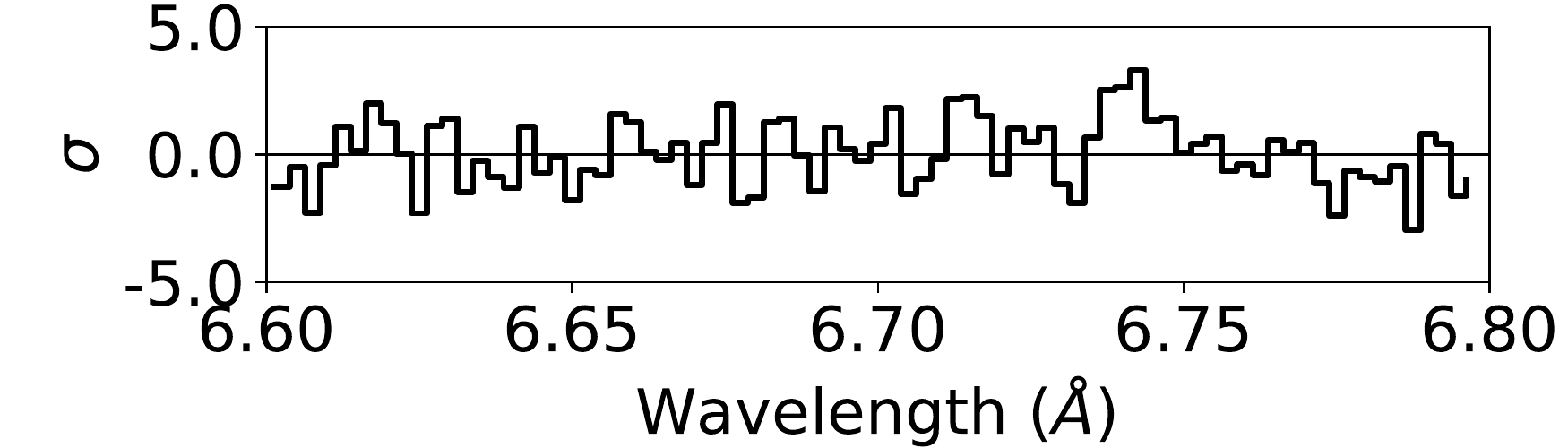}
\includegraphics[scale=0.25]{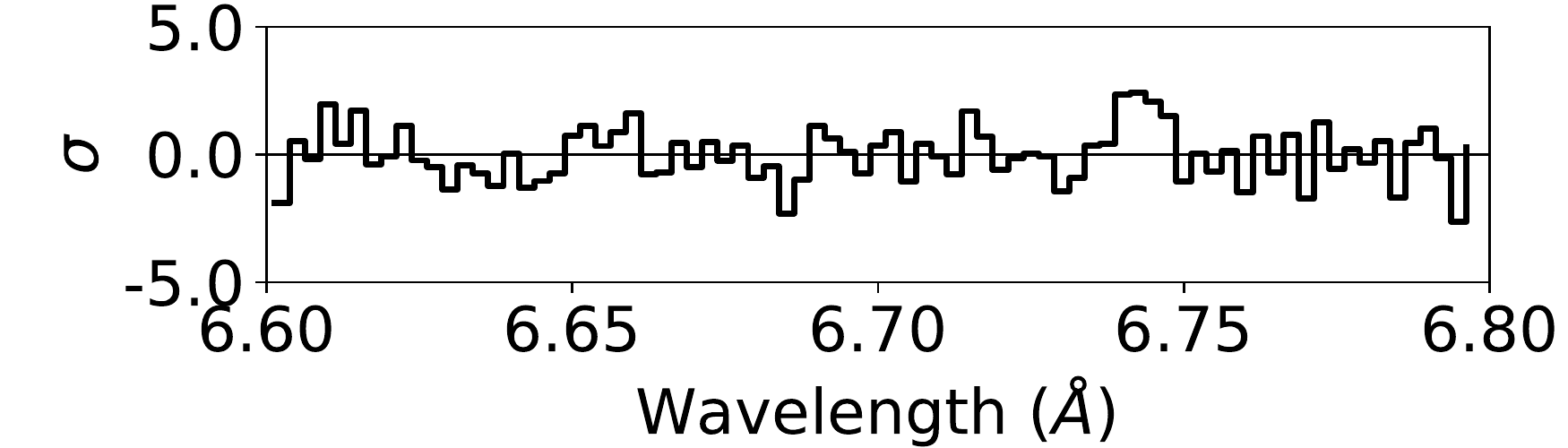}
\includegraphics[scale=0.25]{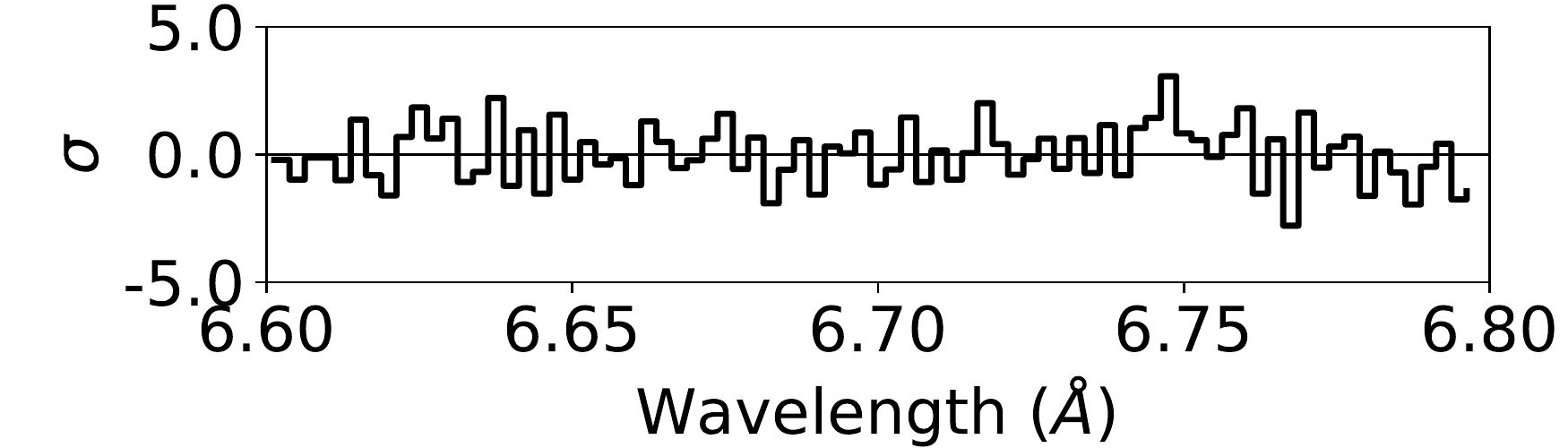}
\includegraphics[scale=0.25]{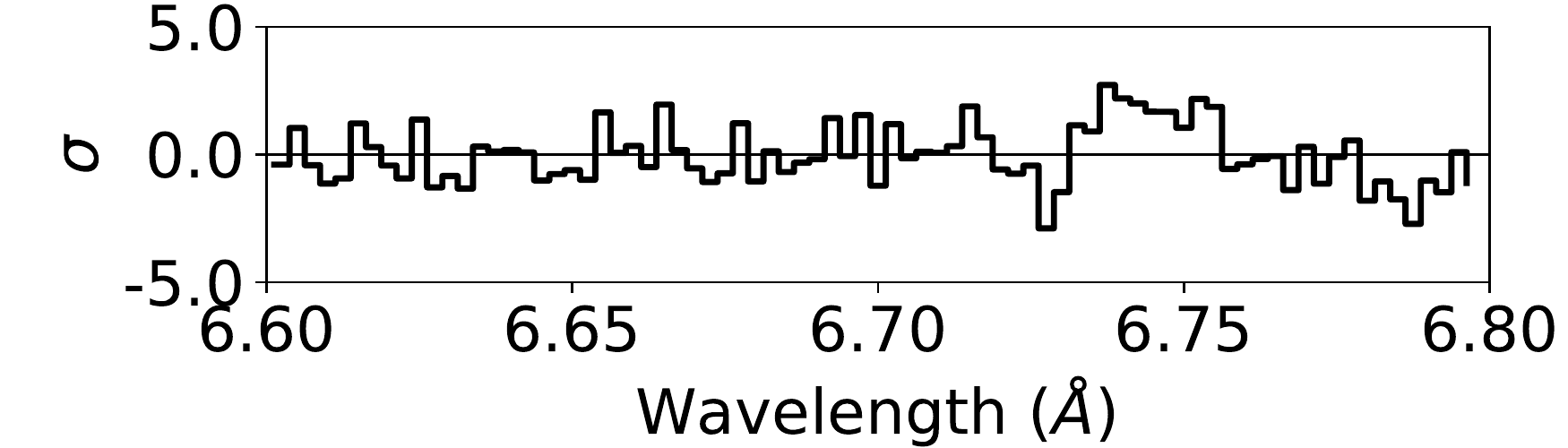}\\
        \hspace*{-5mm}
\includegraphics[scale=0.25]{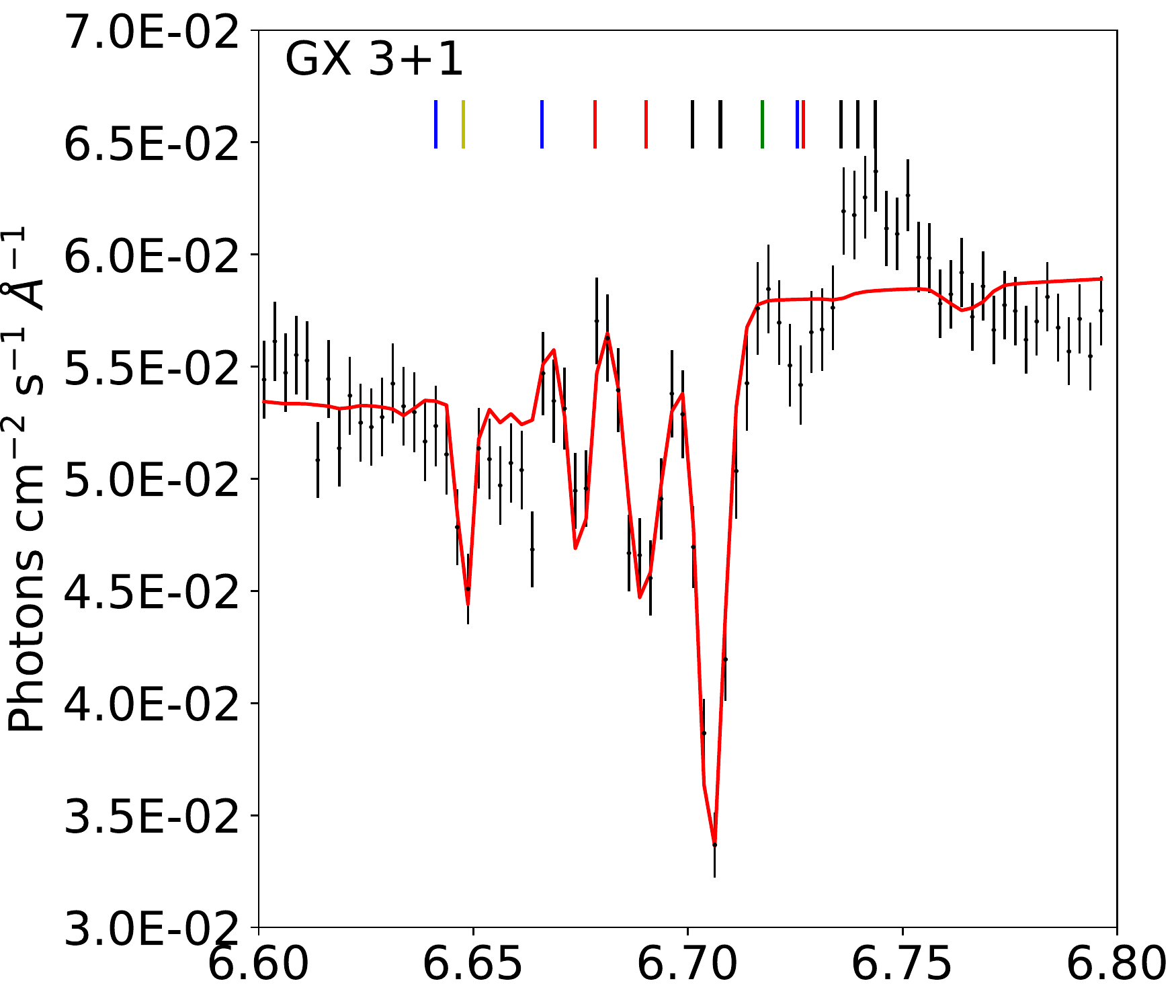}
\includegraphics[scale=0.25]{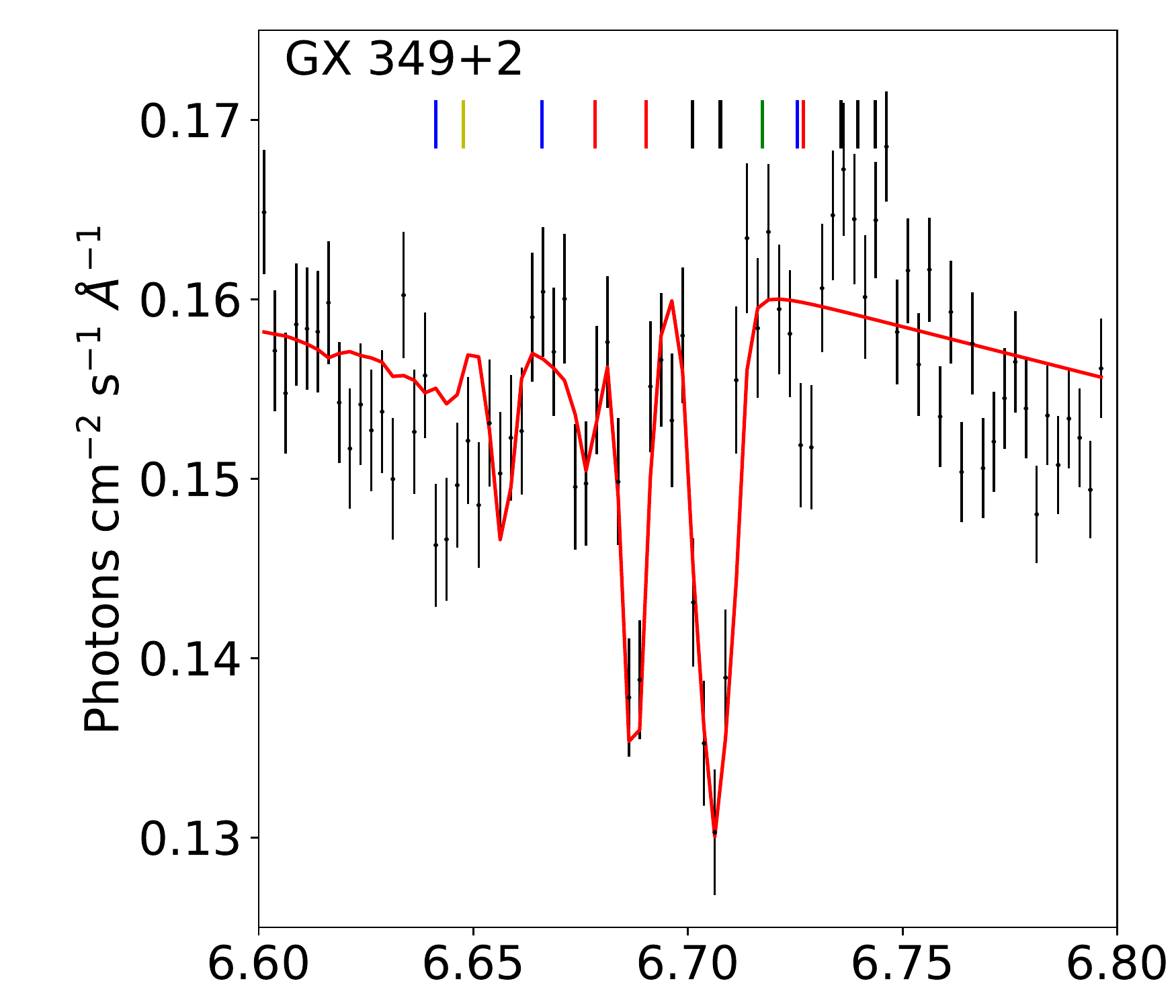}
\includegraphics[scale=0.25]{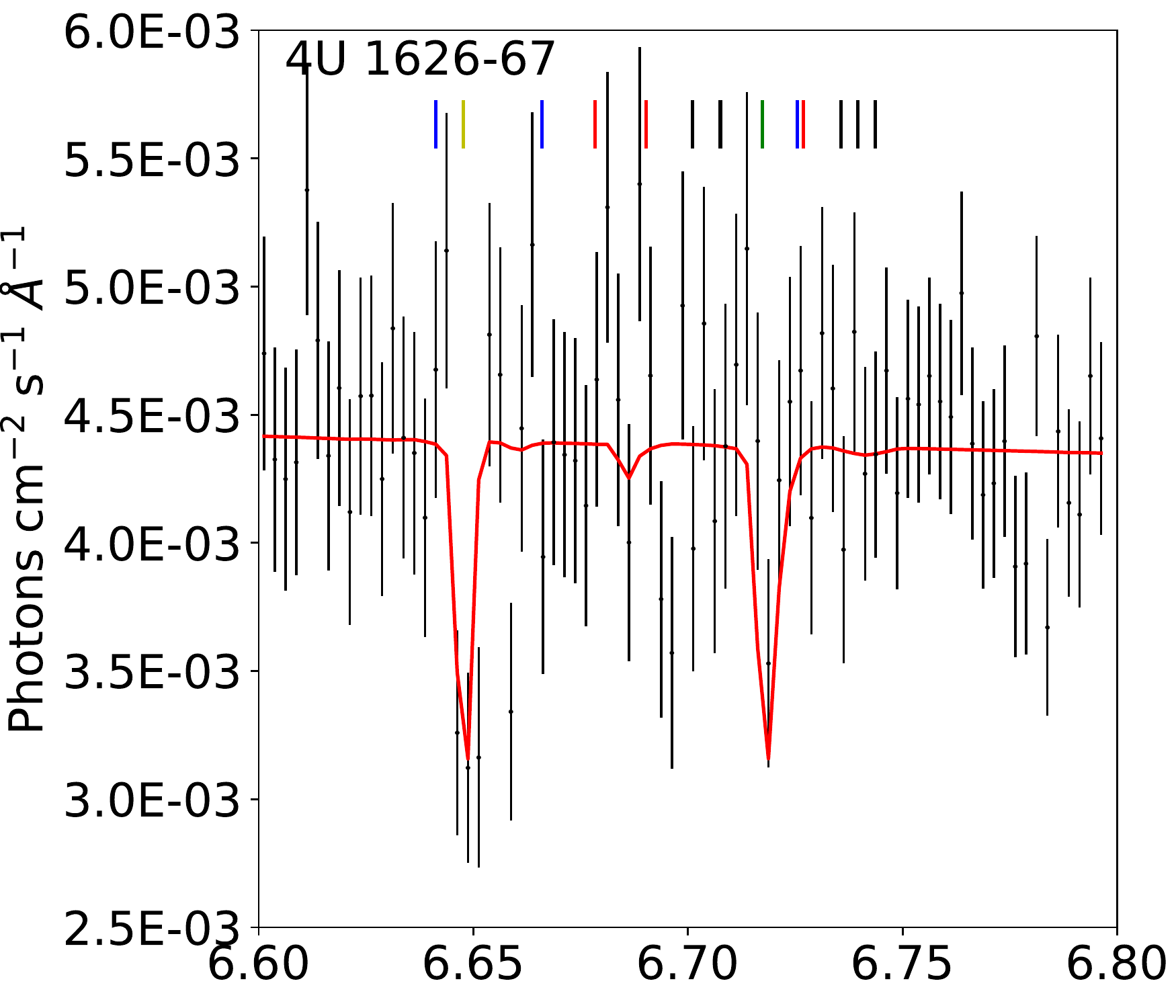}
\includegraphics[scale=0.25]{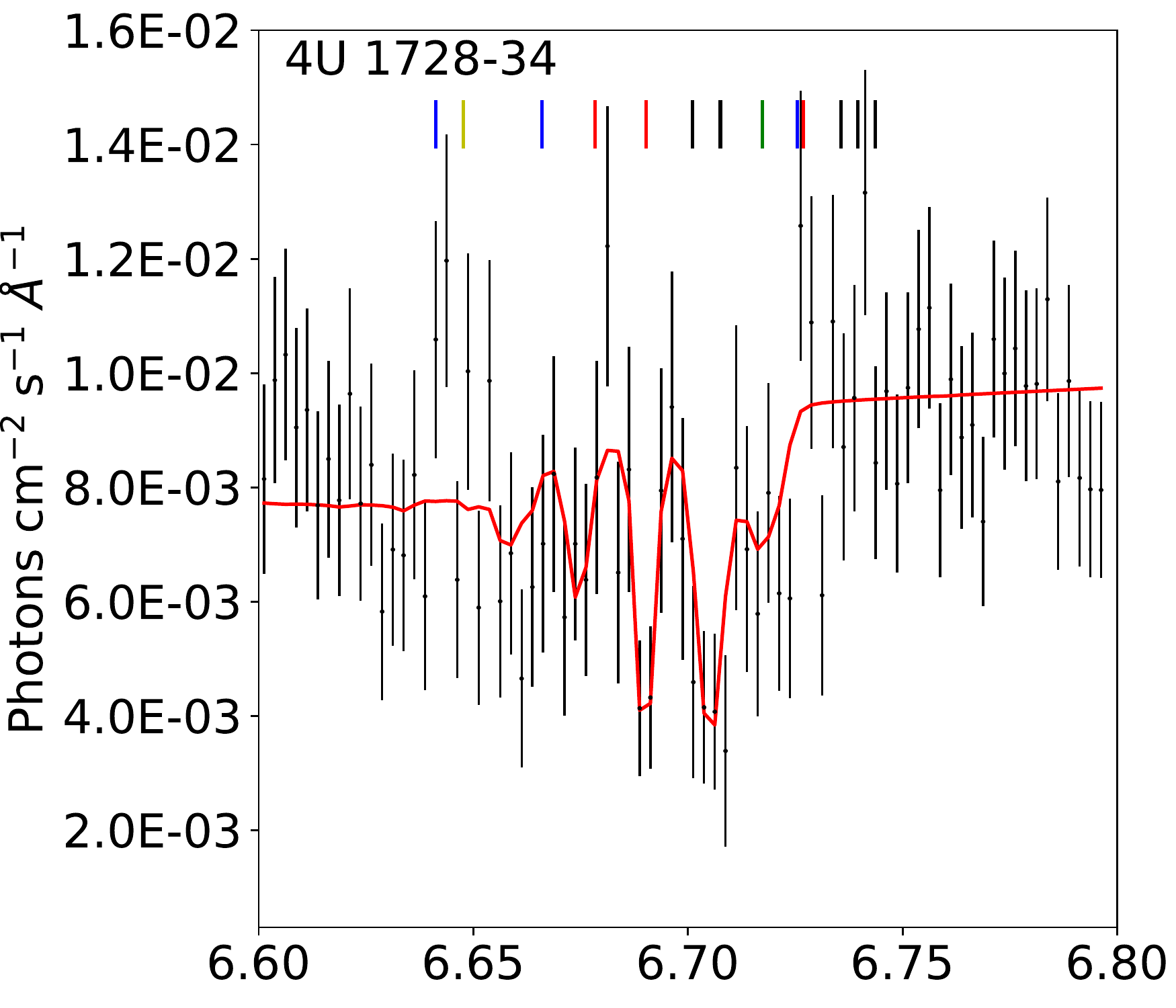}\\
 \hspace*{-5mm}
\includegraphics[scale=0.25]{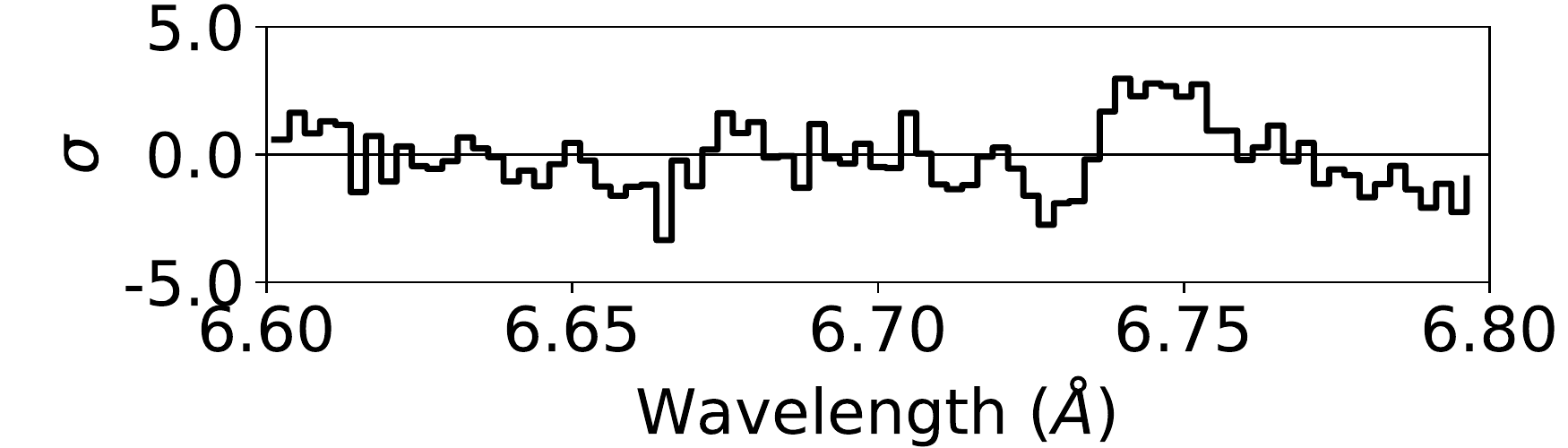}
\includegraphics[scale=0.25]{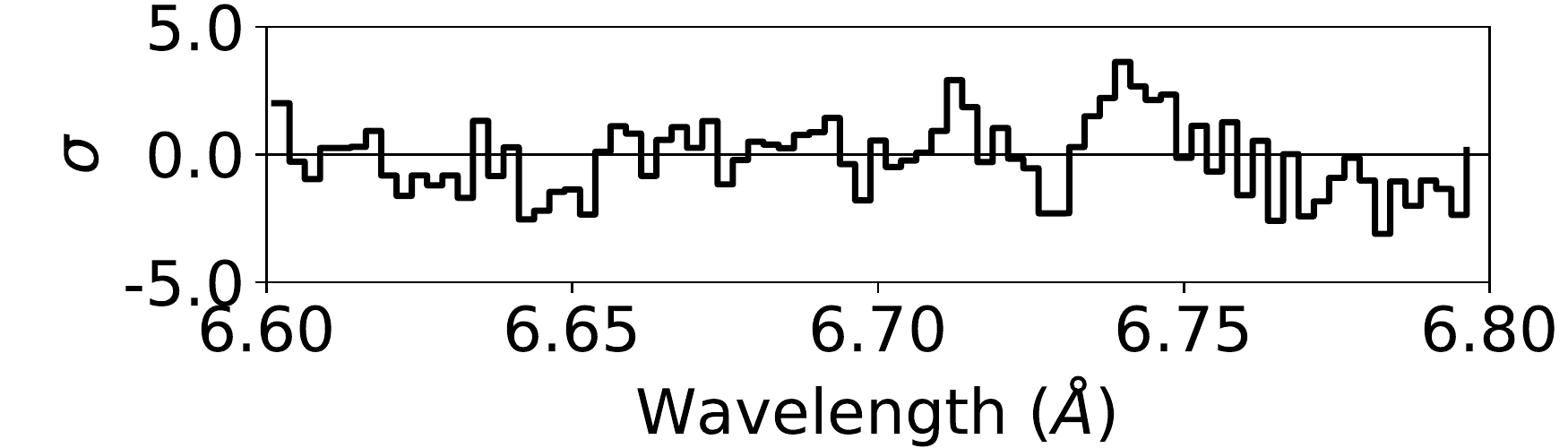}
\includegraphics[scale=0.25]{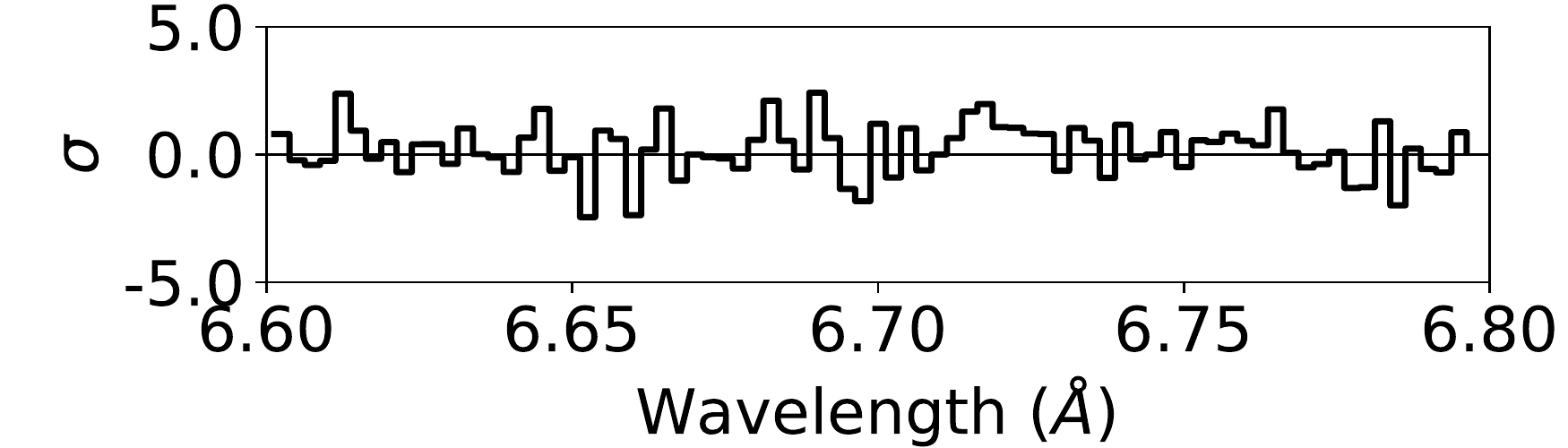}
\includegraphics[scale=0.25]{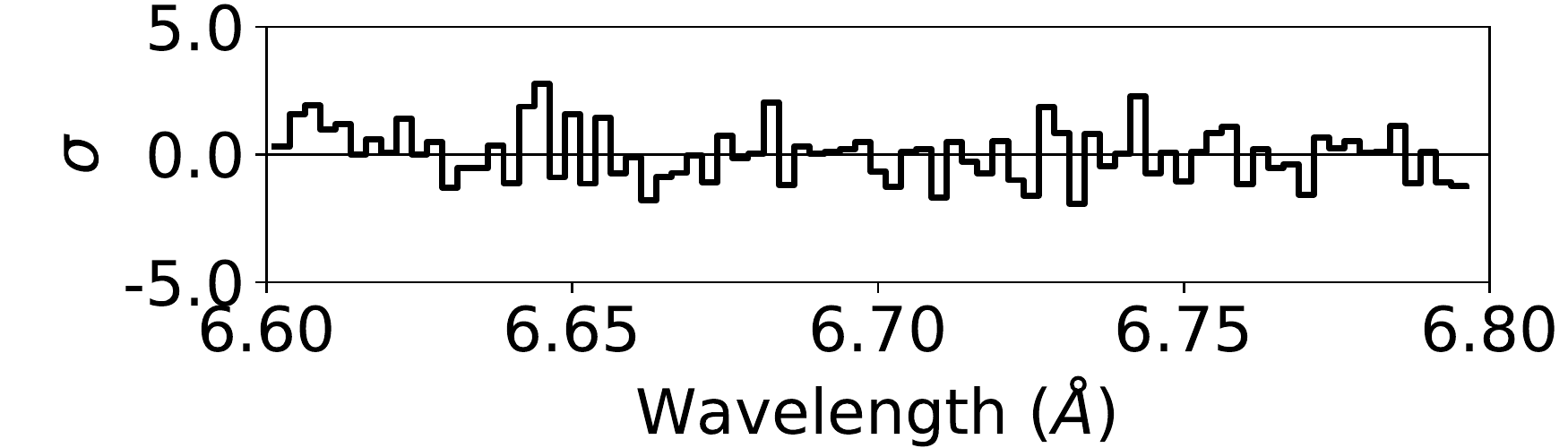}\\
        \hspace*{-5mm}
\includegraphics[scale=0.25]{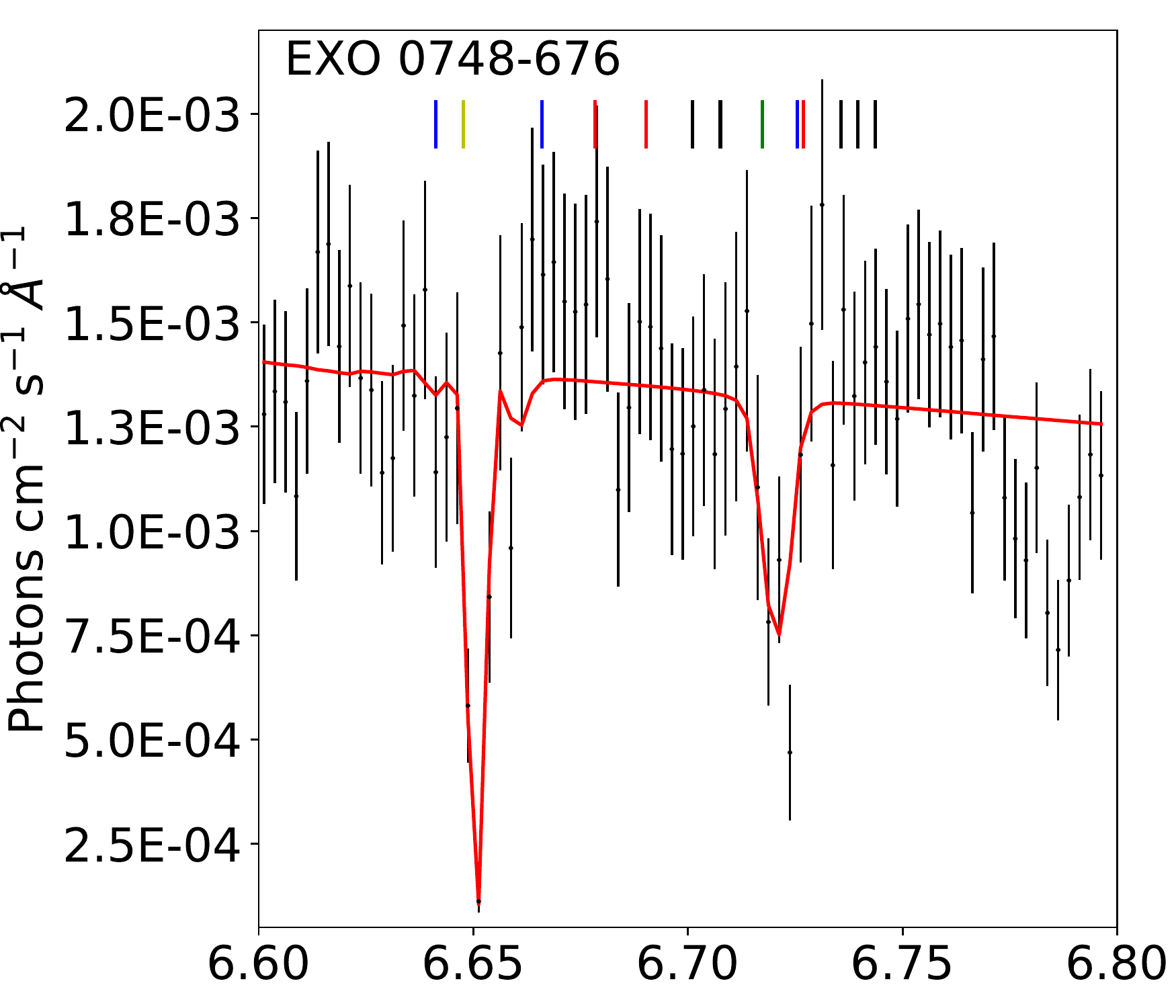}
\includegraphics[scale=0.25]{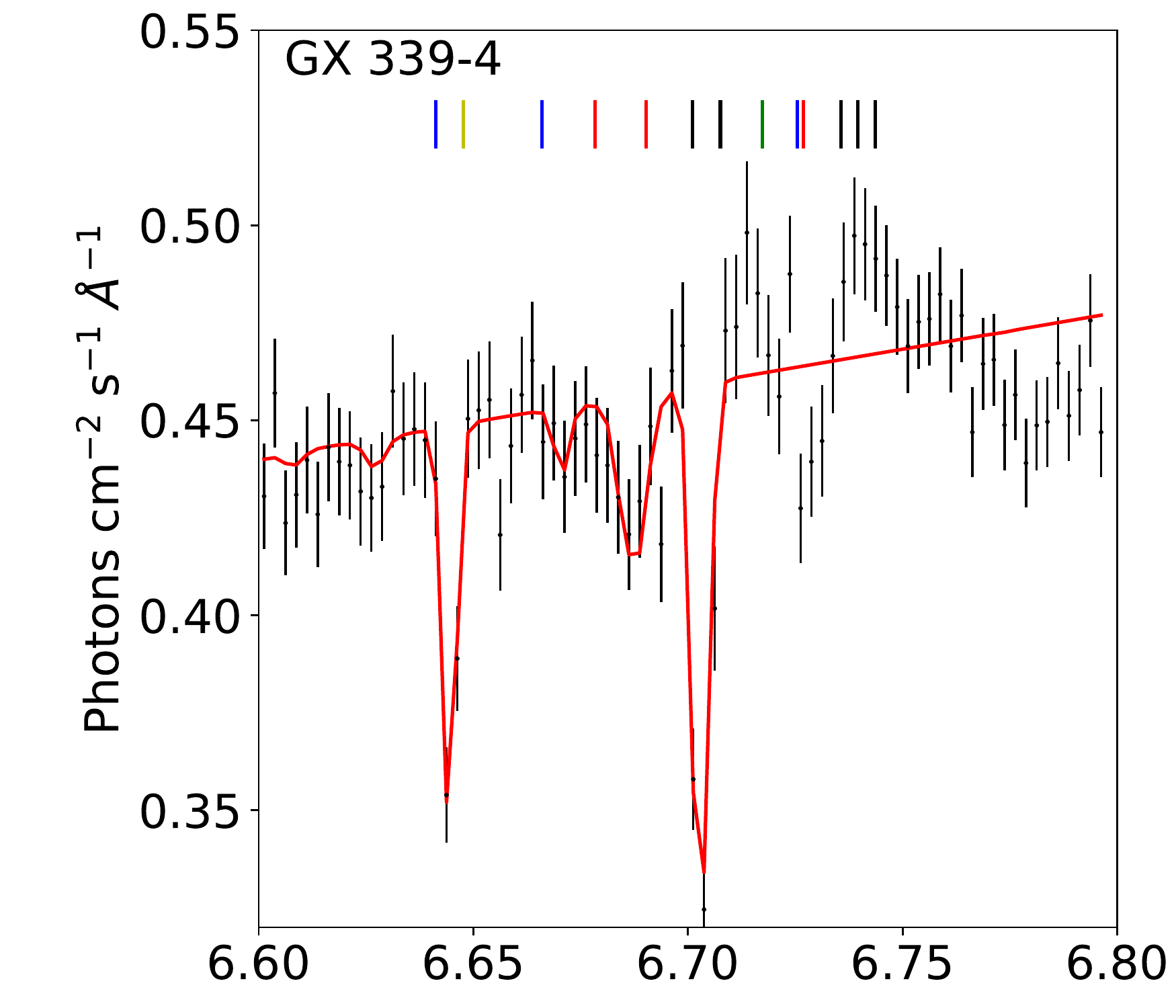}
\includegraphics[scale=0.25]{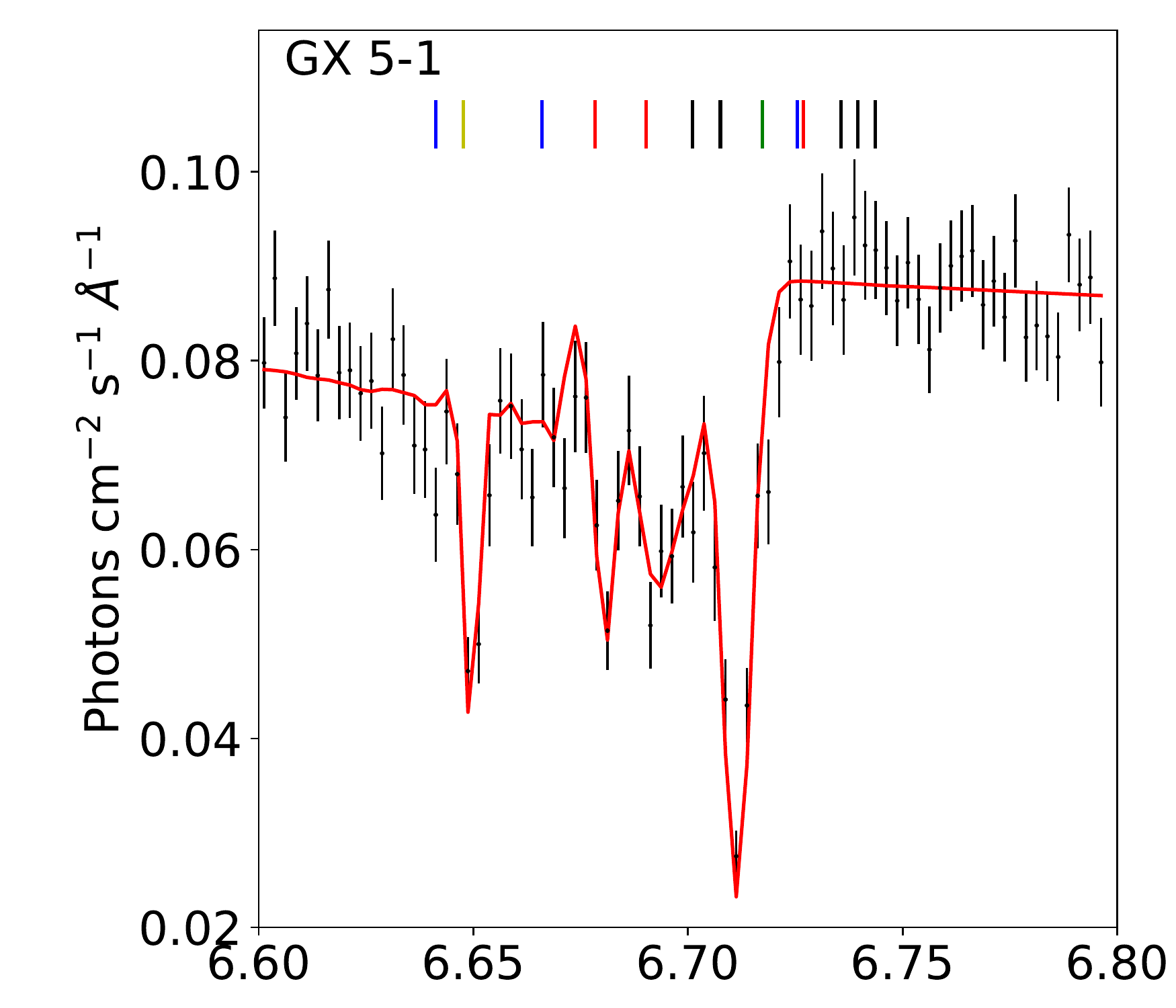}
\includegraphics[scale=0.25]{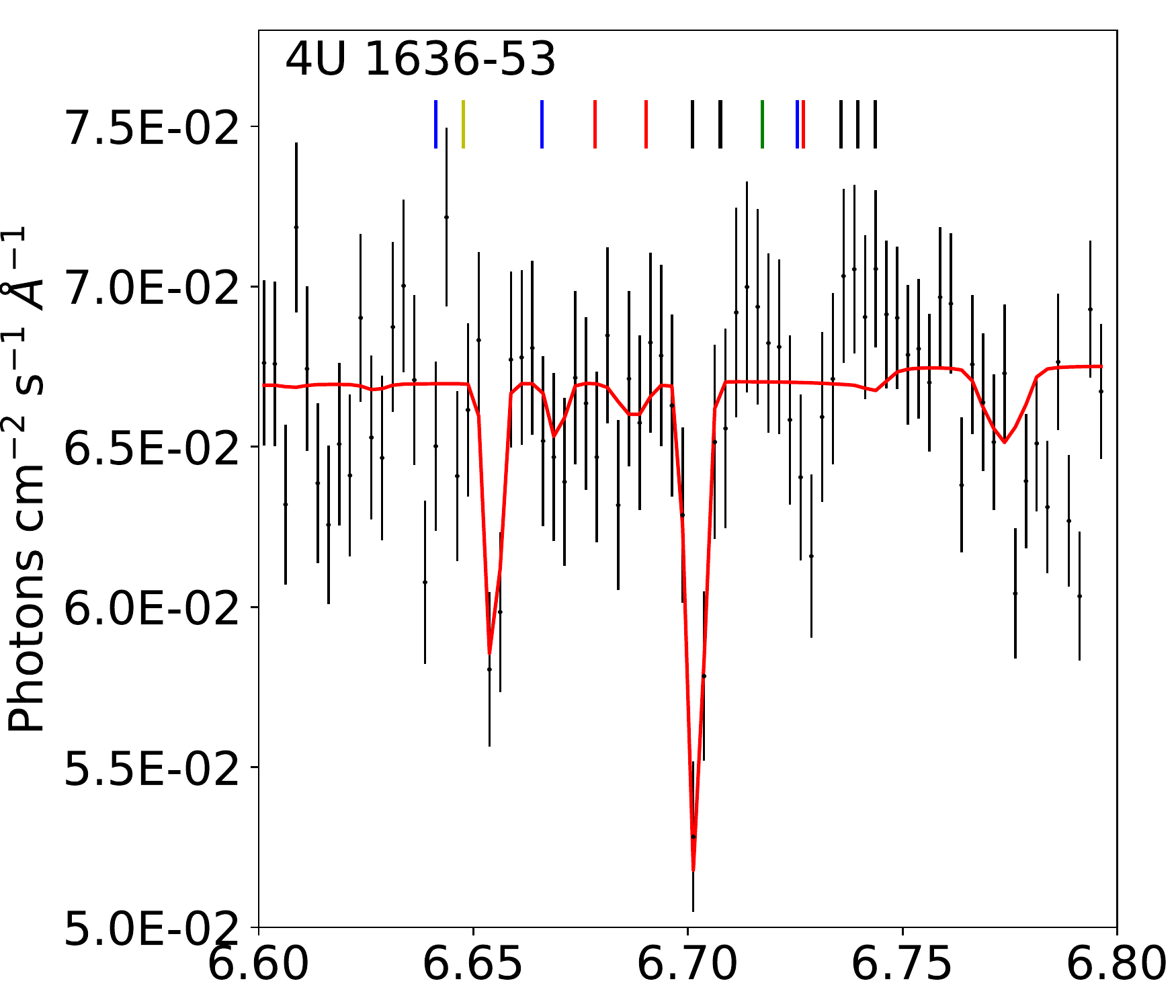}\\
 \hspace*{-5mm}
\includegraphics[scale=0.25]{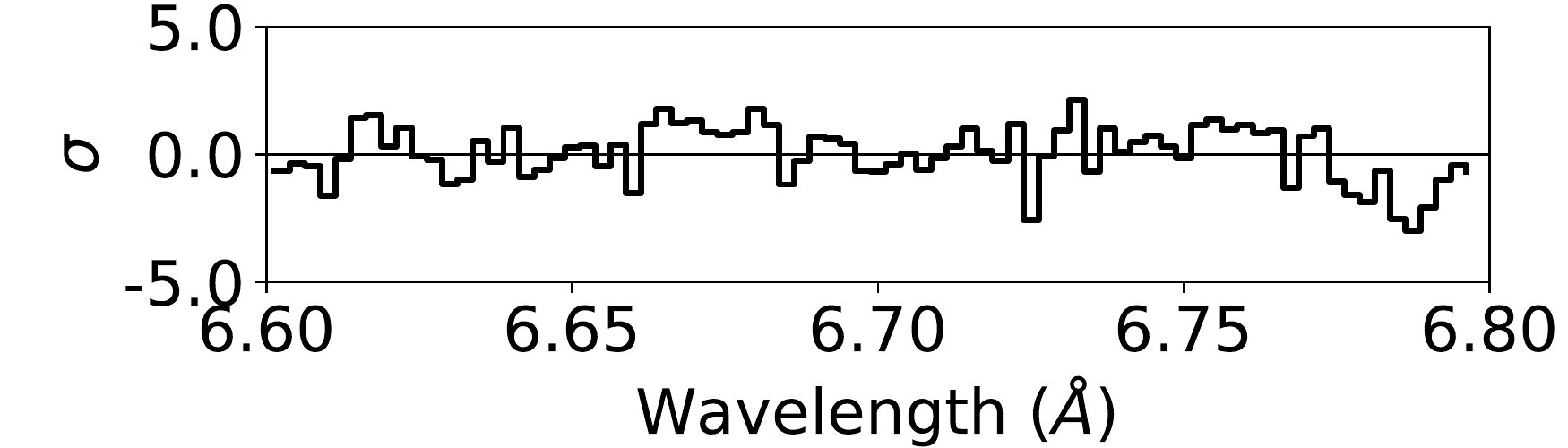}
\includegraphics[scale=0.25]{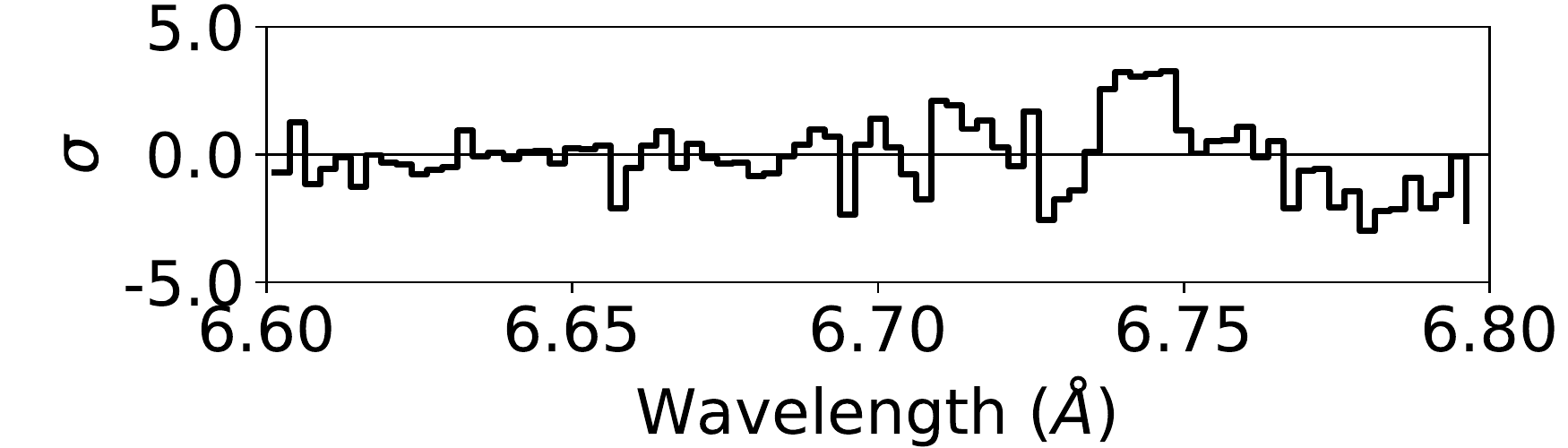}
\includegraphics[scale=0.25]{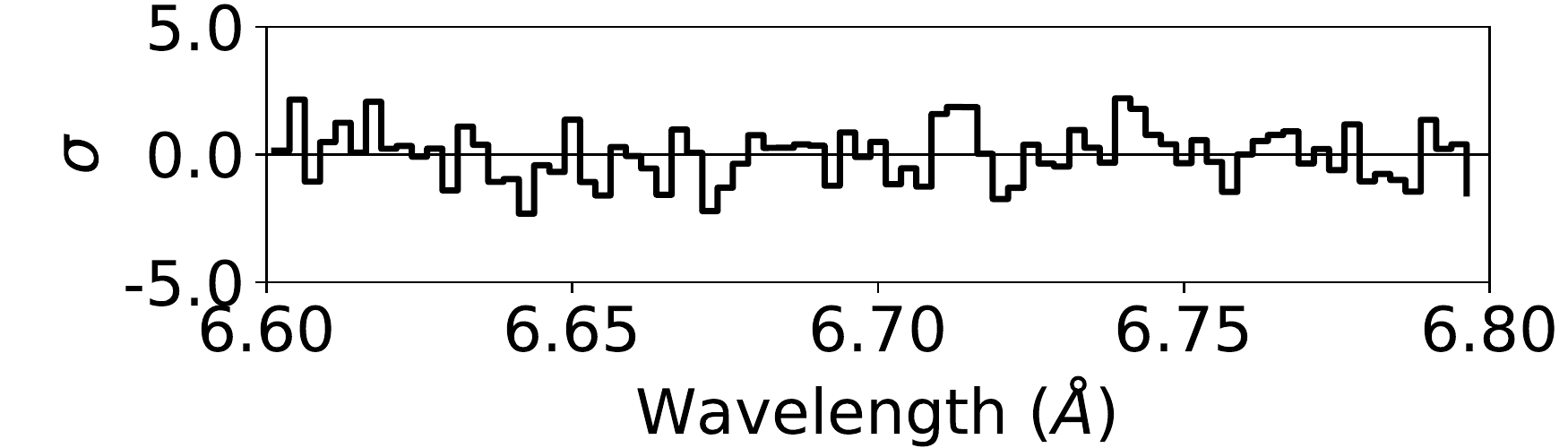}
\includegraphics[scale=0.25]{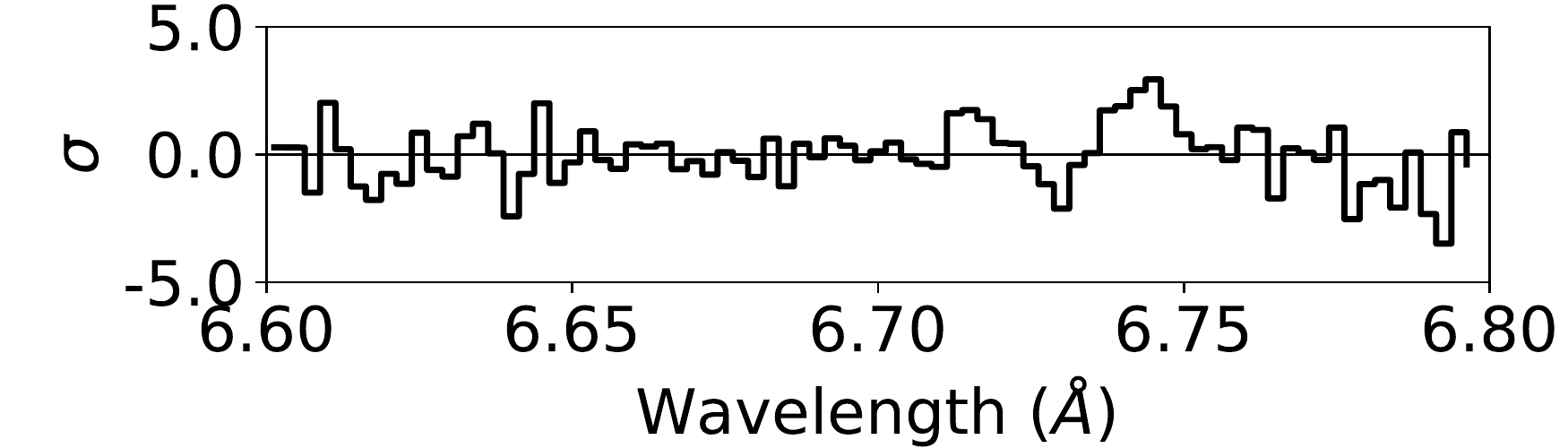}\\
        \hspace*{-5mm}
\includegraphics[scale=0.25]{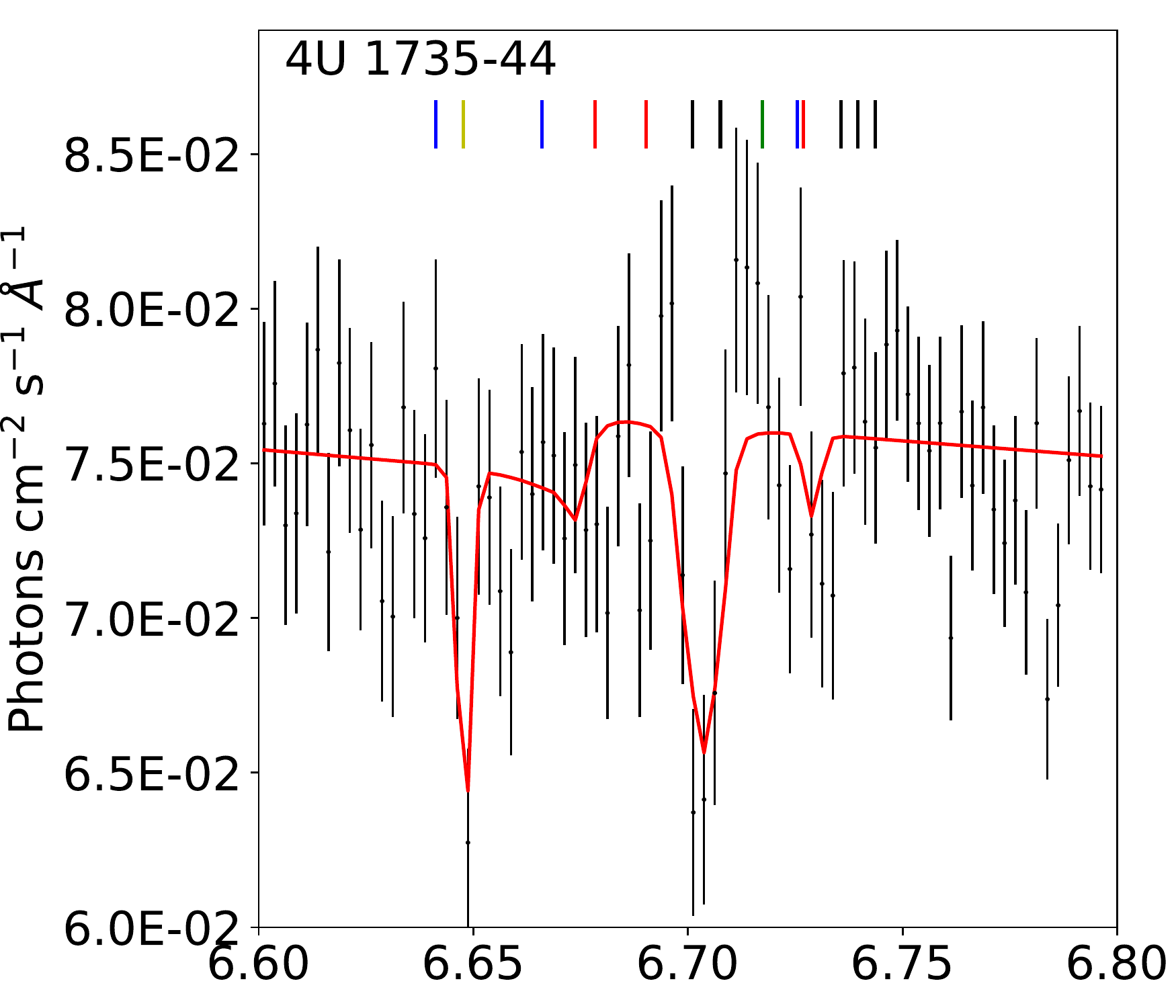}
\includegraphics[scale=0.25]{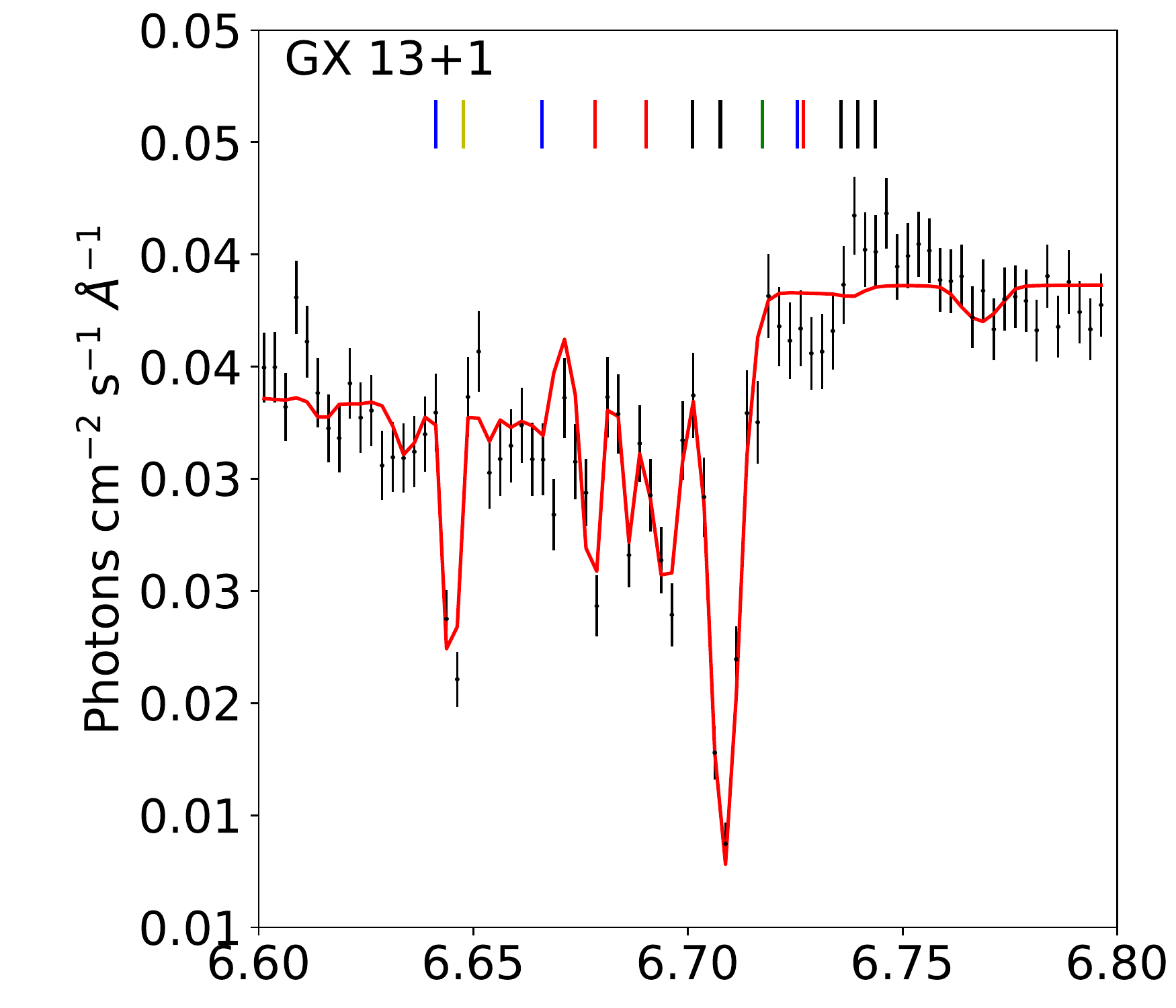}
\includegraphics[scale=0.25]{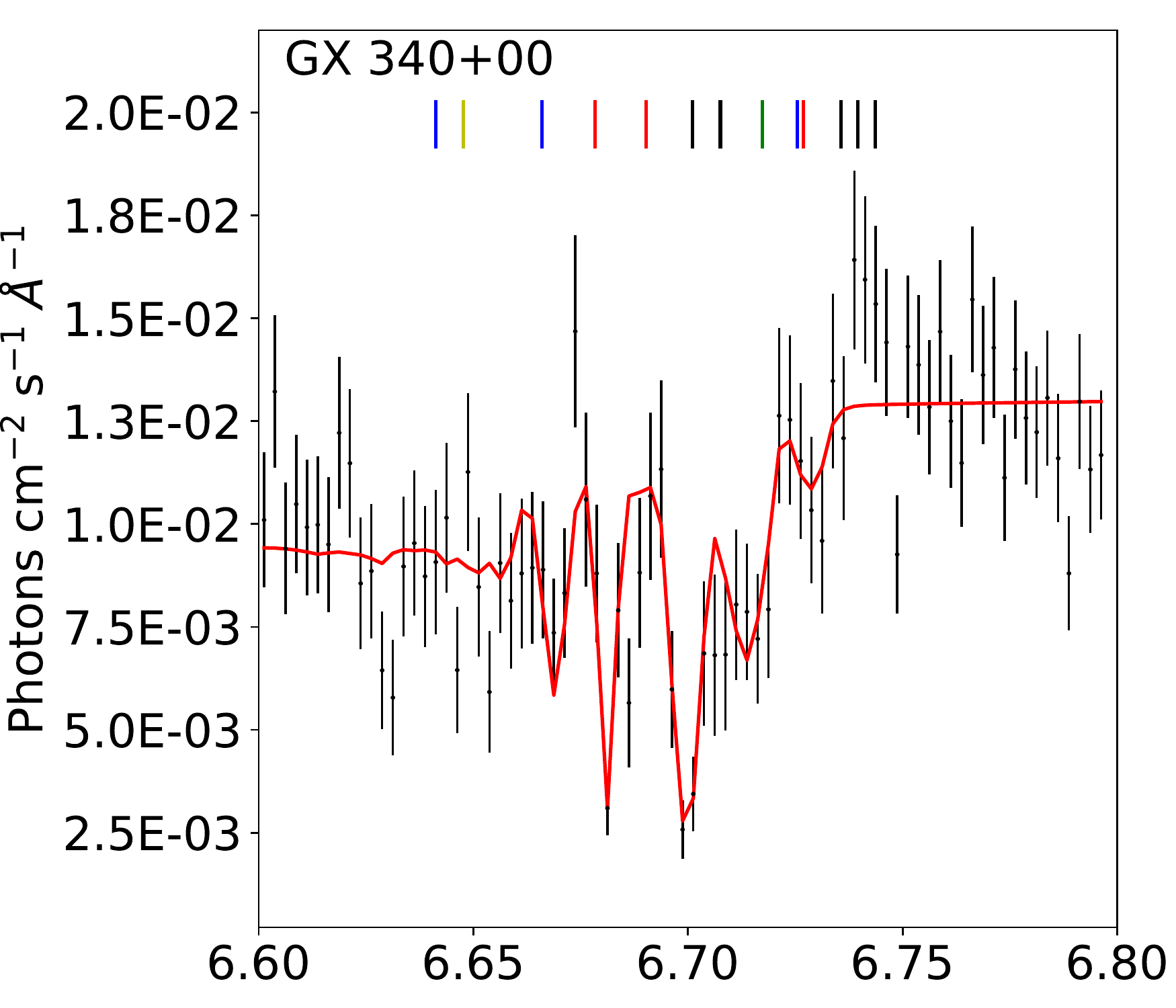}
\includegraphics[scale=0.25]{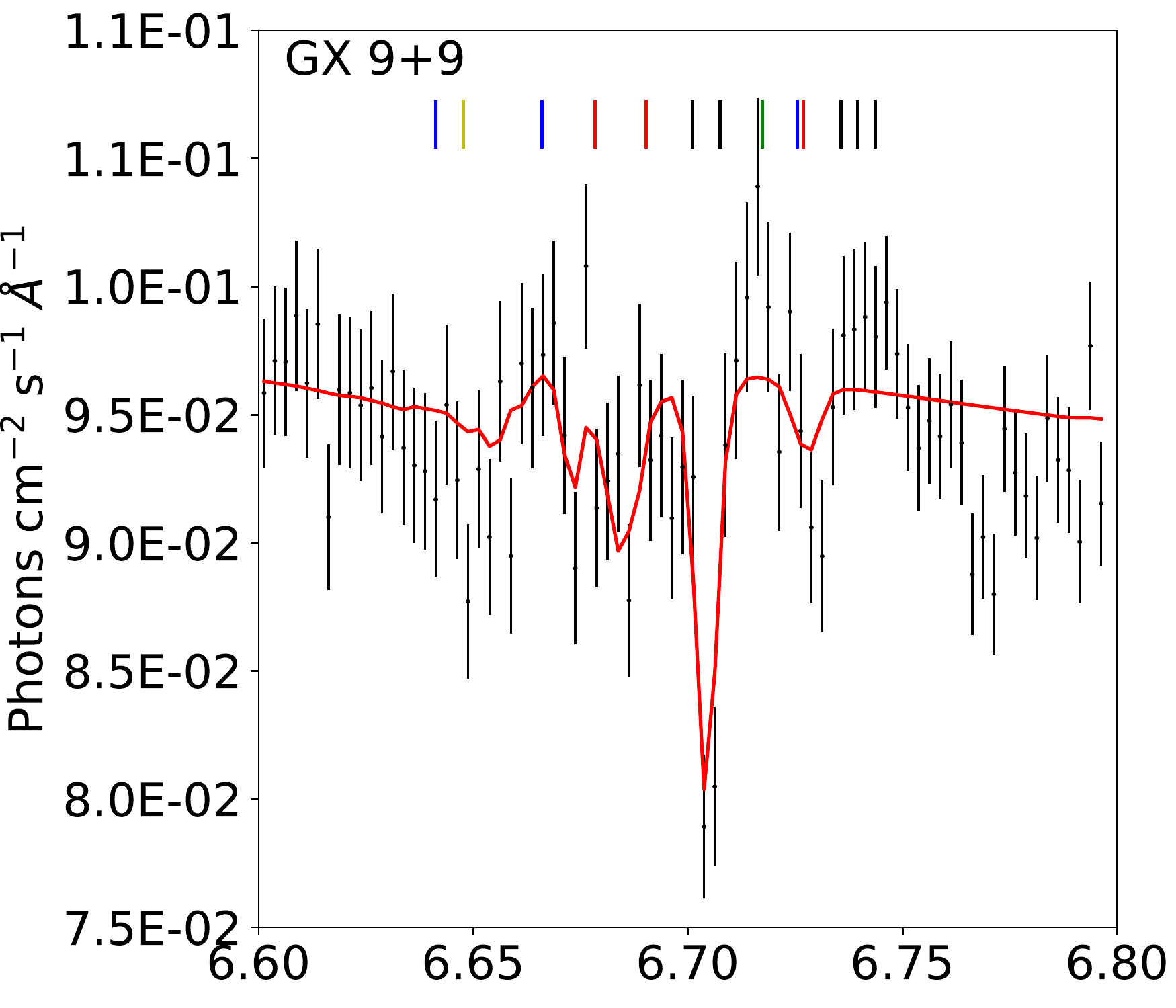}\\
 \hspace*{-5mm}
\includegraphics[scale=0.25]{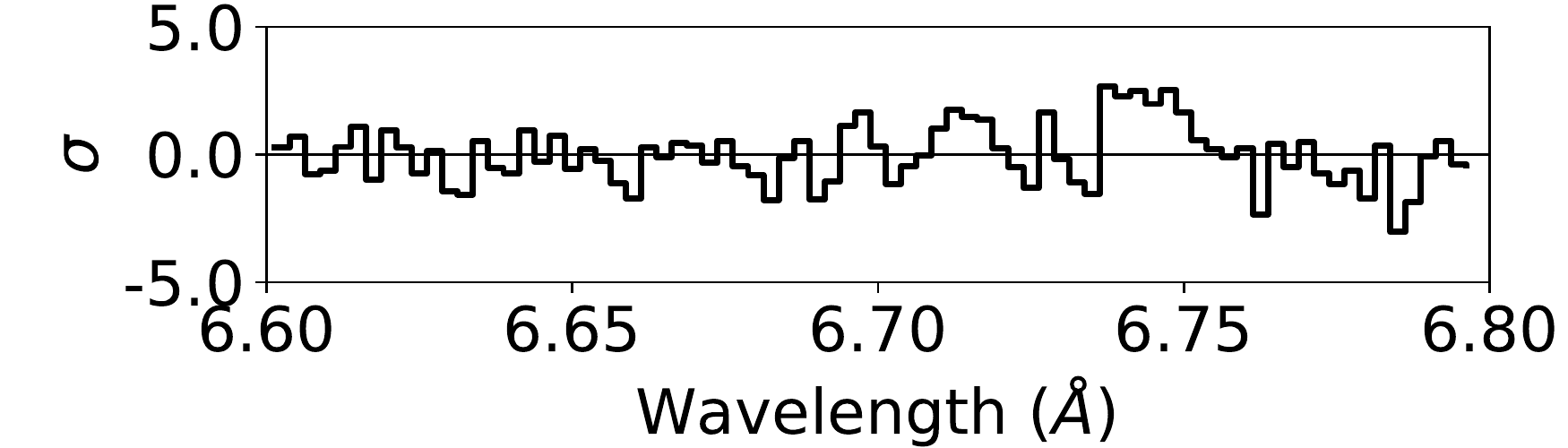}
\includegraphics[scale=0.25]{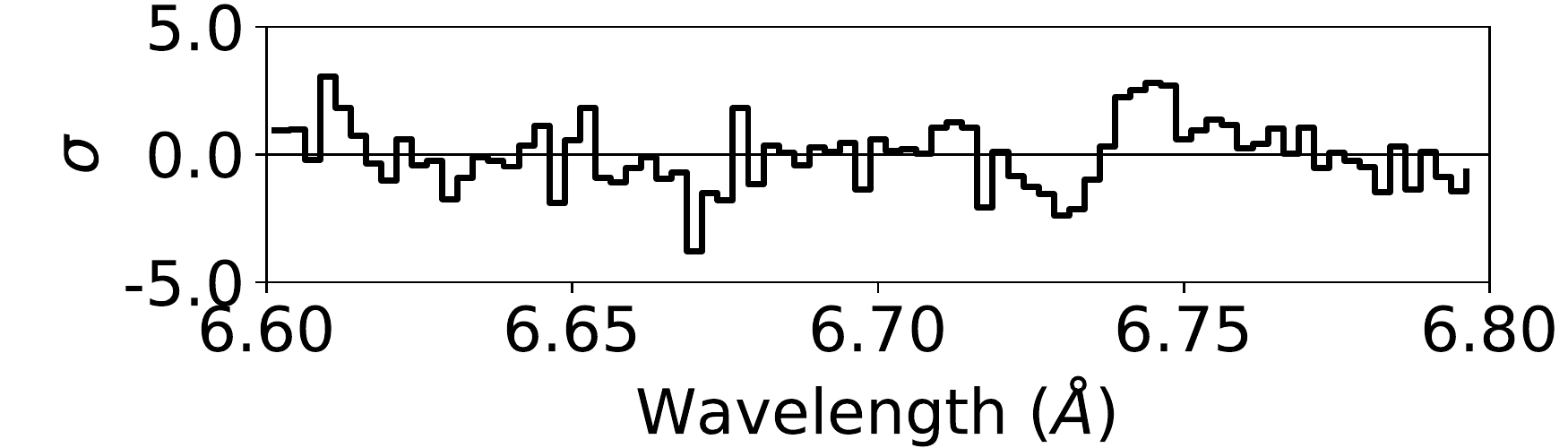}
\includegraphics[scale=0.25]{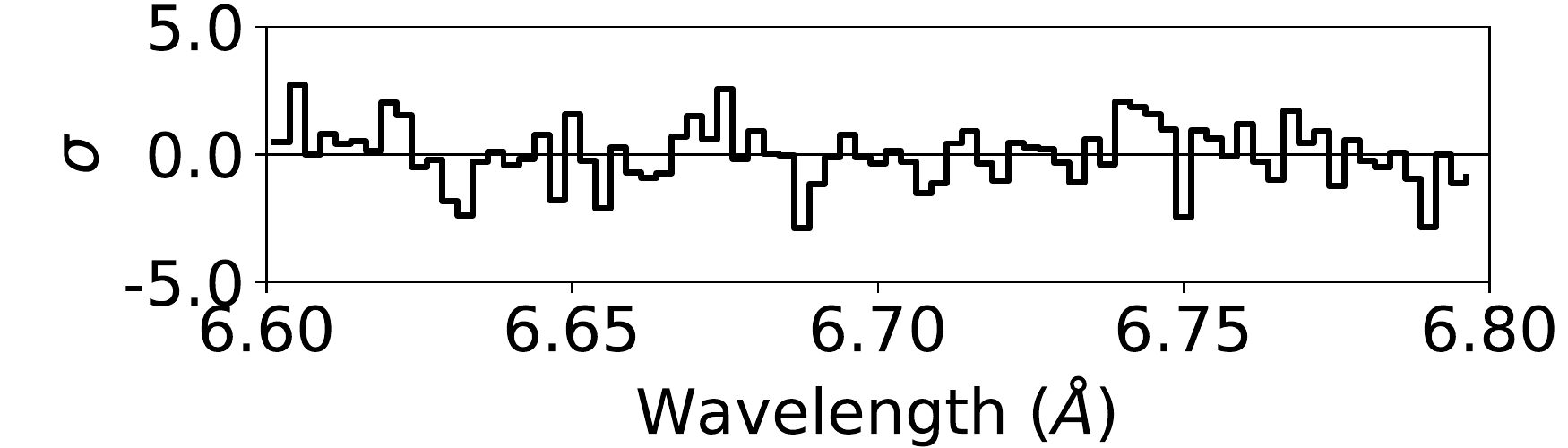}
\includegraphics[scale=0.25]{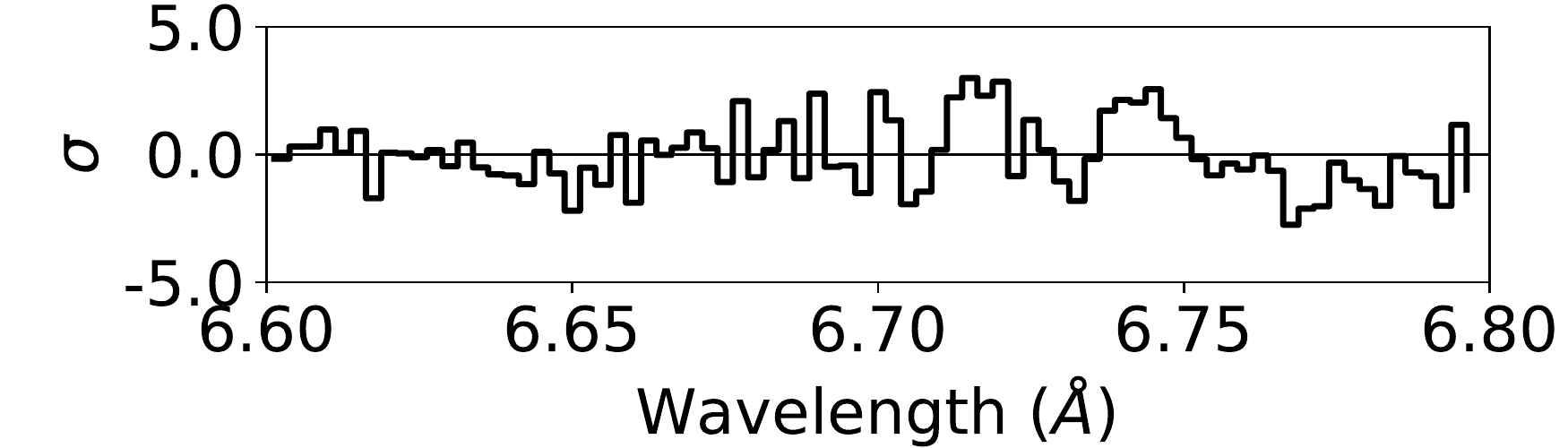}\\
      \caption{Best fit results using {\it Chandra} HETG data for the {\rm Si} K-edge wavelength region. For each source the observations have been combined for illustrative purposes. The main resonances for the {\rm Si}~{\sc i} (black), {\rm Si}~{\sc ii} (red), {\rm Si}~{\sc iii} (blue), {\rm Si}~{\sc xii} (green) and  {\rm Si}~{\sc xiii} (yellow) are indicated. }\label{fig_data_fits}
   \end{figure*}

Figure~\ref{fig_data_fits} shows the best-fit spectra for the individual sources; for each source all observations were combined for illustrative purposes, and residuals are also included. Table~\ref{tab_ismabs} tabulates the best-fit results obtained for each source. Due to the quality of the spectra analyzed most of the values obtained are upper limits. Interestingly, we have found acceptable fits to the observed silicon K-edge spectra with a model consisting of atomic silicon ions similar to the oxygen K-edge absorption region analysis of \citet{gat13a, gat13b, gat14}.

\begin{figure*}
          \centering
\includegraphics[scale=0.45]{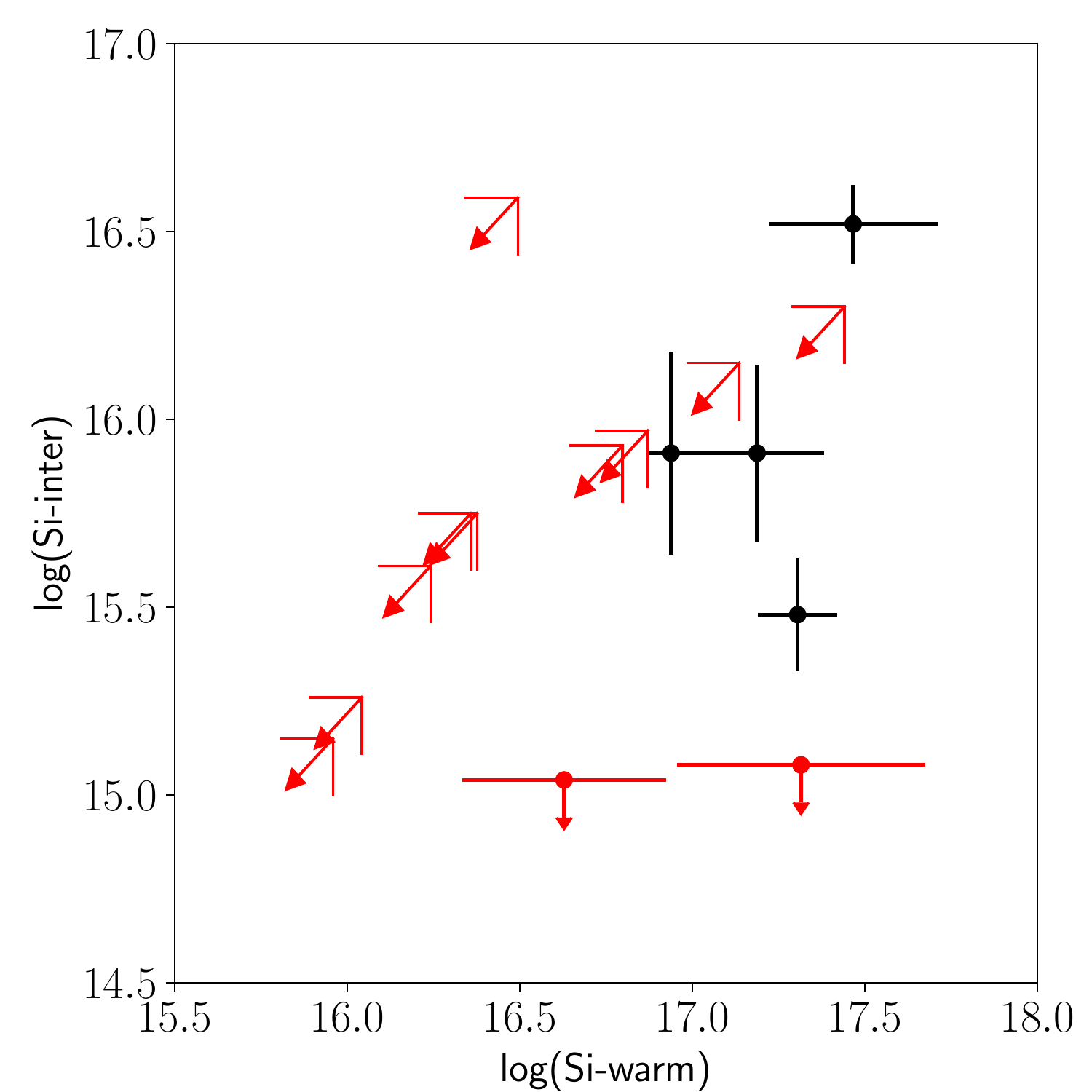}
\includegraphics[scale=0.45]{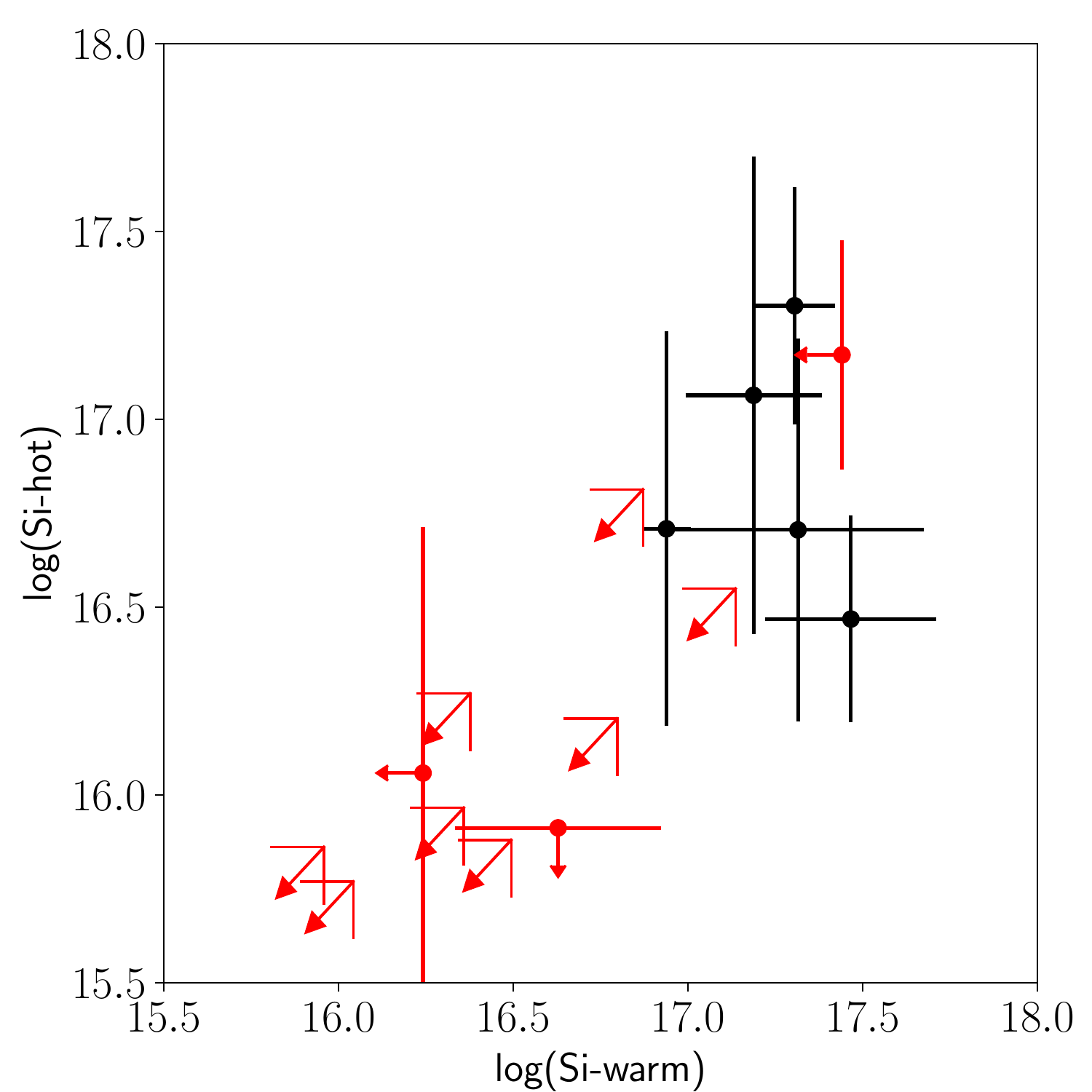}
      \caption{Ratio of intermediate temperature to warm gas column densities (left panel) and ratio of hot to warm gas column densities (right panel) from the individual best fits. Si$_\mathrm{warm}$ corresponds to ({\rm Si}~{\sc i}+{\rm Si}~{\sc ii}),  Si$_\mathrm{inter}$ (i.e. intermediate temperature)  corresponds to {\rm Si}~{\sc iii} and Si$_\mathrm{hot}$ corresponds to ({\rm Si}~{\sc xii}+{\rm Si}~{\sc xiii}). We note that column densities corresponding to the warm and intermediate temperature phases are similar although most of the column densities correspond to upper values.}\label{fig_data_fractions}
   \end{figure*}

Although the solid absorption analysis is beyond the scope of this work, we emphasise that the gaseous component must be carefully modeled before measuring the solid absorption features to estimate accurately the depletion factor of atomic silicon. It is noteworthy that for {\rm Si}~{\sc i} the X-ray dust scattering, which is not included in the current model, would affect the abundance determination by a factor of $\sim$ 2 \citep{cor16,zee19}. Figure \ref{fig_data_fractions} shows the column densities obtained from the best fit. We distinguish between the Si$_\mathrm{warm}$ ({\rm Si}~{\sc i}+{\rm Si}~{\sc ii}), Si$_\mathrm{inter}$ (intermediate temperature phase including {\rm Si}~{\sc iii}) and Si$_\mathrm{hot}$ ({\rm Si}~{\sc xii}+{\rm Si}~{\sc xiii}). We have found that the column densities of the warm and intermediate temperature phases are similar.

Since the physical state of the plasma determines the ionic fractions, the temperature associated to the warm and intermediate temperature phases, assuming collisional ionization equilibrium, results in similar ionic fractions. {\rm Si}~{\sc ii} is also expected to be a tracer of the warm gas given its low photoionization potential \citep{sav96}. For {\rm Si}~{\sc iii}, \citet{col09} obtained column densities $>\log{12.5}$ with a sky-covering fraction of 80--90\% in agreement with our upper limits. Under collisional ionization equilibrium conditions, this ion traces transition regions between the warm and hot gas phases due to shocks \citep{sla15,chi16}, turbulent mixing layers \citep{sav17,she18} and evaporating clouds \citep{dwe08,fox10}. {\rm Si}~{\sc xii} and {\rm Si}~{\sc xiii}, on the other hand, trace hot gas temperatures ($\sim 4{-}5\times 10^{6}$~K). We found that such hot gas does not fill a large fraction of the volume of the Galactic disk \citep{jou06,bre12}. Moreover, \citet{gat18b} have found a similar result for carbon whereby {\rm C}~{\sc ii} dominates over {\rm C}~{\sc i}. An alternative explanation would be the increase of warm material due to the absorber material in the close vicinity of the X-ray binaries \citep{sch16}.  Future analyses, using a complete gas$+$dust model, will help to distinguish the origin of the Si absorbers.

 \begin{table*}
\caption{\label{tab_ismabs}ISM silicon column densities obtained with the {\it ISMabs} model. }
\centering
\begin{tabular}{lccccccccc}
\hline
Source  &  Si\,{\sc i} &  Si\,{\sc ii}  &  Si\,{\sc iii} &  Si\,{\sc xii} &  Si\,{\sc xiii}    &{\tt cash}/d.of.\\
 \hline
\hline
\\
4U~0614+091  &$<1.28$&$ <0.46$&$<0.41$&$<0.47 $&$0.68_{-0.58}^{+1.38} $&  3236/3177 \\
4U~1626-67   &$<7.05$&$<6.60$&$<1.42$&$<0.72$&$<2.85$& 4207/3973  \\
4U~1636-53   &$<0.93$&$<2.20$&$<3.87$&$<0.27$&$<0.49$& 3538/3177  \\
4U~1705-44   &$11.89\pm 0.89$&$3.35_{-1.18}^{+1.23}$&$0.82\pm 0.40 $&$0.88\pm 0.36$&$10.70_{-7.90}^{+11.50}$& 1315/2381  \\
4U~1728-34 	 &$19.29\pm 3.78$&$9.76\pm 3.32$&$3.29\pm 0.75$&$1.31\pm 0.28$&$1.64\pm 0.64 $& 1208/2381  \\
4U~1735-44   &$<0.76$&$<1.64$&$<0.56$&$<0.58$&$<1.29$& 2496/2381  \\
4U~1820-30   &$<2.86$&$<3.38$&$<0.86$&$<0.53$&$<1.07$& 1914/1585  \\
Cygnus~X-2   &$<0.19$&$<0.92$&$<0.18$&$<0.12$&$<0.47$& 4580/3177  \\
EXO~0748-676 &$<3.92$&$<3.53$&$<0.93$&$<4.47$&$2.03_{-0.53}^{+0.32}$& 2738/2381  \\
GX~339-4     &$<0.88$&$<1.42$&$<0.56$&$<0.46$&$<0.47$& 6983/4769  \\
GX~349+2     &$<0.27$&$4.02_{-2.38}^{+2.47}$&$<0.11$&$<0.14$&$<0.68$&  7694/7953  \\
GX~9+9       &$<0.37$&$<0.54$&$<0.14$&$<0.15$&$<0.58$& 1783/1585  \\
GX~340+00    &$<11.71 $&$<15.91$&$<1.98$&$<1.67$&$13.1_{-4.20}^{+5.11}$&  1557/2381  \\
GX~5-1       &$11.56_{-1.26}^{+1.32}$&$8.68\pm 1.33$&$0.30\pm 0.10$&$4.55\pm 0.42$&$15.59_{-8.04}^{+11.11}$& 1068/1585  \\
GX~3+1       &$3.29\pm 0.24$&$5.33\pm 0.48$&$0.82\pm 0.45$&$0.10\pm 0.06$&$5.01_{-2.19}^{+2.86}$& 6604/5565  \\
GX~13+1      &$12.95_{-7.58}^{+9.96}$&$7.79_{-0.77}^{+1.03}$&$<0.12$&$0.82\pm 0.60$&$4.29_{-0.92}^{+1.22}$ &4191/3973  \\
\hline
\multicolumn{5}{l}{ Column densities in units of $10^{16}$cm$^{-2}$.}
 \end{tabular}
\end{table*}

\section{Conclusions}\label{sec_con}

We have analyzed the silicon edge absorption region (6$-$7 \AA) using high-resolution X-ray spectra from 16 LMXBs. First, we combine all spectra to perform a benchmark of the silicon photoabsorption cross sections. We found that the absorption features identified in the spectra agree with the theoretical atomic data, even though the individual resonances of the K$\alpha$ triplet cannot be resolved. Using the {\it ISMabs} model, we have estimated ionic column densities corresponding to the warm, intermediate temperature and hot phases of the gaseous ISM. Our findings highlight the need for accurate modeling of the gaseous component before attempting to address the solid component. Including the solid components in the model will be the next step of the current study. Future X-ray observatories such as {\it XRISM}, {\it Lynx} and {\it Athena} will display the spectral energy resolution necessary to allow an accurate benchmark of the Si K-edge atomic data.

\subsection*{Data availability}
Observations analyzed in this article are available in the Chandra Data Archive (\url{https://cxc.harvard.edu/cda/})

\bibliographystyle{mnras}

\begin{thebibliography}{}
\makeatletter
\relax
\def\mn@urlcharsother{\let\do\@makeother \do\$\do\&\do\#\do\^\do\_\do\%\do\~}
\def\mn@doi{\begingroup\mn@urlcharsother \@ifnextchar [ {\mn@doi@}
  {\mn@doi@[]}}
\def\mn@doi@[#1]#2{\def\@tempa{#1}\ifx\@tempa\@empty \href
  {http://dx.doi.org/#2} {doi:#2}\else \href {http://dx.doi.org/#2} {#1}\fi
  \endgroup}
\def\mn@eprint#1#2{\mn@eprint@#1:#2::\@nil}
\def\mn@eprint@arXiv#1{\href {http://arxiv.org/abs/#1} {{\tt arXiv:#1}}}
\def\mn@eprint@dblp#1{\href {http://dblp.uni-trier.de/rec/bibtex/#1.xml}
  {dblp:#1}}
\def\mn@eprint@#1:#2:#3:#4\@nil{\def\@tempa {#1}\def\@tempb {#2}\def\@tempc
  {#3}\ifx \@tempc \@empty \let \@tempc \@tempb \let \@tempb \@tempa \fi \ifx
  \@tempb \@empty \def\@tempb {arXiv}\fi \@ifundefined
  {mn@eprint@\@tempb}{\@tempb:\@tempc}{\expandafter \expandafter \csname
  mn@eprint@\@tempb\endcsname \expandafter{\@tempc}}}

\bibitem[\protect\citeauthoryear{{Apai} \& {Lauretta}}{{Apai} \&
  {Lauretta}}{2010}]{apa10}
{Apai} D.~A.,  {Lauretta} D.~S.,  2010, {Protoplanetary Dust: Astrophysical and
  Cosmochemical Perspectives}

\bibitem[\protect\citeauthoryear{{Bigiel}, {Leroy}, {Walter}, {Brinks}, {de
  Blok}, {Madore}  \& {Thornley}}{{Bigiel} et~al.}{2008}]{big08}
{Bigiel} F.,  {Leroy} A.,  {Walter} F.,  {Brinks} E.,  {de Blok} W.~J.~G.,
  {Madore} B.,   {Thornley} M.~D.,  2008, \mn@doi [\aj]
  {10.1088/0004-6256/136/6/2846}, \href
  {https://ui.adsabs.harvard.edu/abs/2008AJ....136.2846B} {136, 2846}

\bibitem[\protect\citeauthoryear{{Breitschwerdt}, {de Avillez}, {Feige}  \&
  {Dettbarn}}{{Breitschwerdt} et~al.}{2012}]{bre12}
{Breitschwerdt} D.,  {de Avillez} M.~A.,  {Feige} J.,   {Dettbarn} C.,  2012,
  \mn@doi [Astronomische Nachrichten] {10.1002/asna.201211692}, \href
  {https://ui.adsabs.harvard.edu/abs/2012AN....333..486B} {333, 486}

\bibitem[\protect\citeauthoryear{{Cash}}{{Cash}}{1979}]{cas79}
{Cash} W.,  1979, \mn@doi [\apj] {10.1086/156922}, \href
  {http://adsabs.harvard.edu/abs/1979ApJ...228..939C} {228, 939}

\bibitem[\protect\citeauthoryear{{Chisholm}, {Tremonti}, {Leitherer}, {Chen}
  \& {Wofford}}{{Chisholm} et~al.}{2016}]{chi16}
{Chisholm} J.,  {Tremonti} C.~A.,  {Leitherer} C.,  {Chen} Y.,   {Wofford} A.,
  2016, \mn@doi [\mnras] {10.1093/mnras/stw178}, \href
  {https://ui.adsabs.harvard.edu/abs/2016MNRAS.457.3133C} {457, 3133}

\bibitem[\protect\citeauthoryear{{Christian} \& {Swank}}{{Christian} \&
  {Swank}}{1997}]{chr97}
{Christian} D.~J.,  {Swank} J.~H.,  1997, \mn@doi [\apjs] {10.1086/312970},
  \href {https://ui.adsabs.harvard.edu/abs/1997ApJS..109..177C} {109, 177}

\bibitem[\protect\citeauthoryear{{Collins}, {Shull}  \& {Giroux}}{{Collins}
  et~al.}{2009}]{col09}
{Collins} J.~A.,  {Shull} J.~M.,   {Giroux} M.~L.,  2009, \mn@doi [\apj]
  {10.1088/0004-637X/705/1/962}, \href
  {https://ui.adsabs.harvard.edu/abs/2009ApJ...705..962C} {705, 962}

\bibitem[\protect\citeauthoryear{{Corrales}, {Garc{\'\i}a}, {Wilms}  \&
  {Baganoff}}{{Corrales} et~al.}{2016}]{cor16}
{Corrales} L.~R.,  {Garc{\'\i}a} J.,  {Wilms} J.,   {Baganoff} F.,  2016,
  \mn@doi [\mnras] {10.1093/mnras/stw376}, \href
  {https://ui.adsabs.harvard.edu/abs/2016MNRAS.458.1345C} {458, 1345}

\bibitem[\protect\citeauthoryear{{Draine}}{{Draine}}{2003}]{dra03b}
{Draine} B.~T.,  2003, \mn@doi [\apj] {10.1086/379123}, \href
  {https://ui.adsabs.harvard.edu/abs/2003ApJ...598.1026D} {598, 1026}

\bibitem[\protect\citeauthoryear{{Dwek} et~al.,}{{Dwek} et~al.}{2008}]{dwe08}
{Dwek} E.,  et~al., 2008, \mn@doi [\apj] {10.1086/529038}, \href
  {https://ui.adsabs.harvard.edu/abs/2008ApJ...676.1029D} {676, 1029}

\bibitem[\protect\citeauthoryear{{Fox}, {Wakker}, {Smoker}, {Richter}, {Savage}
   \& {Sembach}}{{Fox} et~al.}{2010}]{fox10}
{Fox} A.~J.,  {Wakker} B.~P.,  {Smoker} J.~V.,  {Richter} P.,  {Savage} B.~D.,
   {Sembach} K.~R.,  2010, \mn@doi [\apj] {10.1088/0004-637X/718/2/1046}, \href
  {https://ui.adsabs.harvard.edu/abs/2010ApJ...718.1046F} {718, 1046}

\bibitem[\protect\citeauthoryear{{Gail}, {Zhukovska}, {Hoppe}  \&
  {Trieloff}}{{Gail} et~al.}{2009}]{gai09}
{Gail} H.~P.,  {Zhukovska} S.~V.,  {Hoppe} P.,   {Trieloff} M.,  2009, \mn@doi
  [\apj] {10.1088/0004-637X/698/2/1136}, \href
  {https://ui.adsabs.harvard.edu/abs/2009ApJ...698.1136G} {698, 1136}

\bibitem[\protect\citeauthoryear{{Galloway}, {Muno}, {Hartman}, {Psaltis}  \&
  {Chakrabarty}}{{Galloway} et~al.}{2008}]{gall08}
{Galloway} D.~K.,  {Muno} M.~P.,  {Hartman} J.~M.,  {Psaltis} D.,
  {Chakrabarty} D.,  2008, \mn@doi [\apjs] {10.1086/592044}, \href
  {http://adsabs.harvard.edu/abs/2008ApJS..179..360G} {179, 360}

\bibitem[\protect\citeauthoryear{{Gatuzz} \& {Churazov}}{{Gatuzz} \&
  {Churazov}}{2018}]{gat18a}
{Gatuzz} E.,  {Churazov} E.,  2018, \mn@doi [\mnras] {10.1093/mnras/stx2776},
  \href {http://adsabs.harvard.edu/abs/2018MNRAS.474..696G} {474, 696}

\bibitem[\protect\citeauthoryear{{Gatuzz} et~al.,}{{Gatuzz}
  et~al.}{2013a}]{gat13a}
{Gatuzz} E.,  et~al., 2013a, \mn@doi [\apj] {10.1088/0004-637X/768/1/60}, \href
  {http://adsabs.harvard.edu/abs/2013ApJ...768...60G} {768, 60}

\bibitem[\protect\citeauthoryear{{Gatuzz} et~al.,}{{Gatuzz}
  et~al.}{2013b}]{gat13b}
{Gatuzz} E.,  et~al., 2013b, \mn@doi [\apj] {10.1088/0004-637X/778/1/83}, \href
  {http://adsabs.harvard.edu/abs/2013ApJ...778...83G} {778, 83}

\bibitem[\protect\citeauthoryear{{Gatuzz}, {Garc{\'{\i}}a}, {Mendoza},
  {Kallman}, {Bautista}  \& {Gorczyca}}{{Gatuzz} et~al.}{2014}]{gat14}
{Gatuzz} E.,  {Garc{\'{\i}}a} J.,  {Mendoza} C.,  {Kallman} T.~R.,  {Bautista}
  M.~A.,   {Gorczyca} T.~W.,  2014, \mn@doi [\apj]
  {10.1088/0004-637X/790/2/131}, \href
  {http://adsabs.harvard.edu/abs/2014ApJ...790..131G} {790, 131}

\bibitem[\protect\citeauthoryear{{Gatuzz}, {Garc{\'{\i}}a}, {Kallman},
  {Mendoza}  \& {Gorczyca}}{{Gatuzz} et~al.}{2015}]{gat15}
{Gatuzz} E.,  {Garc{\'{\i}}a} J.,  {Kallman} T.~R.,  {Mendoza} C.,   {Gorczyca}
  T.~W.,  2015, \mn@doi [\apj] {10.1088/0004-637X/800/1/29}, \href
  {http://adsabs.harvard.edu/abs/2015ApJ...800...29G} {800, 29}

\bibitem[\protect\citeauthoryear{{Gatuzz}, {Garc{\'{\i}}a}, {Kallman}  \&
  {Mendoza}}{{Gatuzz} et~al.}{2016}]{gat16}
{Gatuzz} E.,  {Garc{\'{\i}}a} J.~A.,  {Kallman} T.~R.,   {Mendoza} C.,  2016,
  \mn@doi [\aap] {10.1051/0004-6361/201527752}, \href
  {http://adsabs.harvard.edu/abs/2016A%26A...588A.111G} {588, A111}

\bibitem[\protect\citeauthoryear{Gatuzz, Ness, Gorczyca, Hasoglu, Kallman  \&
  García}{Gatuzz et~al.}{2018}]{gat18b}
Gatuzz E.,  Ness J.-U.,  Gorczyca T.~W.,  Hasoglu M.~F.,  Kallman T.~R.,
  {Garc{\'{\i}}a} J.~A.,  2018, \mn@doi [\mnras] {10.1093/mnras/sty1517}, 479, 2457

\bibitem[\protect\citeauthoryear{{Grimm}, {Gilfanov}  \& {Sunyaev}}{{Grimm}
  et~al.}{2002}]{gri02}
{Grimm} H.-J.,  {Gilfanov} M.,   {Sunyaev} R.,  2002, \mn@doi [\aap]
  {10.1051/0004-6361:20020826}, \href
  {http://cdsads.u-strasbg.fr/abs/2002A%26A...391..923G} {391, 923}

\bibitem[\protect\citeauthoryear{{Hanner} \& {Zolensky}}{{Hanner} \&
  {Zolensky}}{2010}]{han10}
{Hanner} M.~S.,  {Zolensky} M.~E.,  2010, {The Mineralogy of Cometary Dust}.
pp 203--232, \mn@doi{10.1007/978-3-642-13259-9_4}

\bibitem[\protect\citeauthoryear{{Henning}, {Gr{\"u}n}  \&
  {Steinacker}}{{Henning} et~al.}{2009}]{hen09}
{Henning} T.,  {Gr{\"u}n} E.,   {Steinacker} J.,  2009, {Cosmic Dust - Near and
  Far}.
 Vol. 414

\bibitem[\protect\citeauthoryear{{Hynes}, {Steeghs}, {Casares}, {Charles}  \&
  {O'Brien}}{{Hynes} et~al.}{2004}]{hyn04}
{Hynes} R.~I.,  {Steeghs} D.,  {Casares} J.,  {Charles} P.~A.,   {O'Brien} K.,
  2004, \mn@doi [\apj] {10.1086/421014}, \href
  {http://cdsads.u-strasbg.fr/abs/2004ApJ...609..317H} {609, 317}

\bibitem[\protect\citeauthoryear{{Joachimi}, {Gatuzz}, {Garc{\'{\i}}a}  \&
  {Kallman}}{{Joachimi} et~al.}{2016}]{joa16}
{Joachimi} K.,  {Gatuzz} E.,  {Garc{\'{\i}}a} J.~A.,   {Kallman} T.~R.,  2016,
  \mn@doi [\mnras] {10.1093/mnras/stw1371}, \href
  {http://adsabs.harvard.edu/abs/2016MNRAS.461..352J} {461, 352}

\bibitem[\protect\citeauthoryear{{Jonker} \& {Nelemans}}{{Jonker} \&
  {Nelemans}}{2004}]{jon04}
{Jonker} P.~G.,  {Nelemans} G.,  2004, \mn@doi [\mnras]
  {10.1111/j.1365-2966.2004.08193.x}, \href
  {http://cdsads.u-strasbg.fr/abs/2004MNRAS.354..355J} {354, 355}

\bibitem[\protect\citeauthoryear{{Joung} \& {Mac Low}}{{Joung} \& {Mac
  Low}}{2006}]{jou06}
{Joung} M.~K.~R.,  {Mac Low} M.-M.,  2006, \mn@doi [\apj] {10.1086/508795},
  \href {https://ui.adsabs.harvard.edu/abs/2006ApJ...653.1266J} {653, 1266}

\bibitem[\protect\citeauthoryear{{Juett}, {Schulz}  \& {Chakrabarty}}{{Juett}
  et~al.}{2004}]{jue04}
{Juett} A.~M.,  {Schulz} N.~S.,   {Chakrabarty} D.,  2004, \mn@doi [\apj]
  {10.1086/422511}, \href {http://adsabs.harvard.edu/abs/2004ApJ...612..308J}
  {612, 308}

\bibitem[\protect\citeauthoryear{{Juett}, {Schulz}, {Chakrabarty}  \&
  {Gorczyca}}{{Juett} et~al.}{2006}]{jue06}
{Juett} A.~M.,  {Schulz} N.~S.,  {Chakrabarty} D.,   {Gorczyca} T.~W.,  2006,
  \mn@doi [\apj] {10.1086/506189}, \href
  {http://adsabs.harvard.edu/abs/2006ApJ...648.1066J} {648, 1066}

\bibitem[\protect\citeauthoryear{{Kuulkers}, {den Hartog}, {in't Zand},
  {Verbunt}, {Harris}  \& {Cocchi}}{{Kuulkers} et~al.}{2003}]{kul03}
{Kuulkers} E.,  {den Hartog} P.~R.,  {in't Zand} J.~J.~M.,  {Verbunt} F.~W.~M.,
   {Harris} W.~E.,   {Cocchi} M.,  2003, \mn@doi [\aap]
  {10.1051/0004-6361:20021781}, \href
  {http://cdsads.u-strasbg.fr/abs/2003A%26A...399..663K} {399, 663}

\bibitem[\protect\citeauthoryear{{Lada}, {Lombardi}  \& {Alves}}{{Lada}
  et~al.}{2010}]{lad10}
{Lada} C.~J.,  {Lombardi} M.,   {Alves} J.~F.,  2010, \mn@doi [\apj]
  {10.1088/0004-637X/724/1/687}, \href
  {https://ui.adsabs.harvard.edu/abs/2010ApJ...724..687L} {724, 687}

\bibitem[\protect\citeauthoryear{{Leroy}, {Walter}, {Brinks}, {Bigiel}, {de
  Blok}, {Madore}  \& {Thornley}}{{Leroy} et~al.}{2008}]{ler08}
{Leroy} A.~K.,  {Walter} F.,  {Brinks} E.,  {Bigiel} F.,  {de Blok} W.~J.~G.,
  {Madore} B.,   {Thornley} M.~D.,  2008, \mn@doi [\aj]
  {10.1088/0004-6256/136/6/2782}, \href
  {https://ui.adsabs.harvard.edu/abs/2008AJ....136.2782L} {136, 2782}

\bibitem[\protect\citeauthoryear{{Liao}, {Zhang}  \& {Yao}}{{Liao}
  et~al.}{2013}]{lia13}
{Liao} J.-Y.,  {Zhang} S.-N.,   {Yao} Y.,  2013, \mn@doi [\apj]
  {10.1088/0004-637X/774/2/116}, \href
  {http://adsabs.harvard.edu/abs/2013ApJ...774..116L} {774, 116}

\bibitem[\protect\citeauthoryear{{Lilly}, {Carollo}, {Pipino}, {Renzini}  \&
  {Peng}}{{Lilly} et~al.}{2013}]{lil13}
{Lilly} S.~J.,  {Carollo} C.~M.,  {Pipino} A.,  {Renzini} A.,   {Peng} Y.,
  2013, \mn@doi [\apj] {10.1088/0004-637X/772/2/119}, \href
  {https://ui.adsabs.harvard.edu/abs/2013ApJ...772..119L} {772, 119}

\bibitem[\protect\citeauthoryear{{Luo} \& {Fang}}{{Luo} \&
  {Fang}}{2014}]{luo14}
{Luo} Y.,  {Fang} T.,  2014, \mn@doi [\apj] {10.1088/0004-637X/780/2/170},
  \href {http://adsabs.harvard.edu/abs/2014ApJ...780..170L} {780, 170}

\bibitem[\protect\citeauthoryear{{Oosterbroek}, {Barret}, {Guainazzi}  \&
  {Ford}}{{Oosterbroek} et~al.}{2001}]{oos01}
{Oosterbroek} T.,  {Barret} D.,  {Guainazzi} M.,   {Ford} E.~C.,  2001, \mn@doi
  [\aap] {10.1051/0004-6361:20000028}, \href
  {https://ui.adsabs.harvard.edu/abs/2001A&A...366..138O} {366, 138}

\bibitem[\protect\citeauthoryear{{Paerels} et~al.,}{{Paerels}
  et~al.}{2001}]{pae01b}
{Paerels} F.,  et~al., 2001, \mn@doi [\apj] {10.1086/318251}, \href
  {http://cdsads.u-strasbg.fr/abs/2001ApJ...546..338P} {546, 338}

\bibitem[\protect\citeauthoryear{{Pinto}, {Kaastra}, {Costantini}  \& {de
  Vries}}{{Pinto} et~al.}{2013}]{pin13}
{Pinto} C.,  {Kaastra} J.~S.,  {Costantini} E.,   {de Vries} C.,  2013, \mn@doi
  [\aap] {10.1051/0004-6361/201220481}, \href
  {http://adsabs.harvard.edu/abs/2013A%26A...551A..25P} {551, A25}

\bibitem[\protect\citeauthoryear{{Rupke} \& {Veilleux}}{{Rupke} \&
  {Veilleux}}{2013}]{rup13}
{Rupke} D. S.~N.,  {Veilleux} S.,  2013, \mn@doi [\apj]
  {10.1088/0004-637X/768/1/75}, \href
  {https://ui.adsabs.harvard.edu/abs/2013ApJ...768...75R} {768, 75}

\bibitem[\protect\citeauthoryear{{Savage} \& {Sembach}}{{Savage} \&
  {Sembach}}{1996}]{sav96}
{Savage} B.~D.,  {Sembach} K.~R.,  1996, \mn@doi [\araa]
  {10.1146/annurev.astro.34.1.279}, \href
  {https://ui.adsabs.harvard.edu/abs/1996ARA&A..34..279S} {34, 279}

\bibitem[\protect\citeauthoryear{{Savage} et~al.,}{{Savage}
  et~al.}{2017}]{sav17}
{Savage} B.~D.,  et~al., 2017, \mn@doi [\apjs] {10.3847/1538-4365/aa8f4c},
  \href {https://ui.adsabs.harvard.edu/abs/2017ApJS..232...25S} {232, 25}

\bibitem[\protect\citeauthoryear{{Schulz}, {Corrales}  \& {Canizares}}{{Schulz}
  et~al.}{2016}]{sch16}
{Schulz} N.~S.,  {Corrales} L.,   {Canizares} C.~R.,  2016, \mn@doi [\apj]
  {10.3847/0004-637X/827/1/49}, \href
  {https://ui.adsabs.harvard.edu/abs/2016ApJ...827...49S} {827, 49}

\bibitem[\protect\citeauthoryear{{Schulz}, {Chakrabarty}  \&
  {Marshall}}{{Schulz} et~al.}{2019}]{sch19}
{Schulz} N.~S.,  {Chakrabarty} D.,   {Marshall} H.~L.,  2019, arXiv e-prints,
  \href {https://ui.adsabs.harvard.edu/abs/2019arXiv191111684S} {p.
  arXiv:1911.11684}

\bibitem[\protect\citeauthoryear{{Shelton} \& {Kwak}}{{Shelton} \&
  {Kwak}}{2018}]{she18}
{Shelton} R.~L.,  {Kwak} K.,  2018, \mn@doi [\apj] {10.3847/1538-4357/aadced},
  \href {https://ui.adsabs.harvard.edu/abs/2018ApJ...866...34S} {866, 34}

\bibitem[\protect\citeauthoryear{{Slavin}, {Dwek}  \& {Jones}}{{Slavin}
  et~al.}{2015}]{sla15}
{Slavin} J.~D.,  {Dwek} E.,   {Jones} A.~P.,  2015, \mn@doi [\apj]
  {10.1088/0004-637X/803/1/7}, \href
  {https://ui.adsabs.harvard.edu/abs/2015ApJ...803....7S} {803, 7}

\bibitem[\protect\citeauthoryear{{Tonnesen} \& {Bryan}}{{Tonnesen} \&
  {Bryan}}{2009}]{ton09}
{Tonnesen} S.,  {Bryan} G.~L.,  2009, \mn@doi [\apj]
  {10.1088/0004-637X/694/2/789}, \href
  {https://ui.adsabs.harvard.edu/abs/2009ApJ...694..789T} {694, 789}

\bibitem[\protect\citeauthoryear{{Wada} \& {Norman}}{{Wada} \&
  {Norman}}{2001}]{wad01}
{Wada} K.,  {Norman} C.~A.,  2001, \mn@doi [\apj] {10.1086/318344}, \href
  {https://ui.adsabs.harvard.edu/abs/2001ApJ...547..172W} {547, 172}

\bibitem[\protect\citeauthoryear{{Willingale}, {Starling}, {Beardmore},
  {Tanvir}  \& {O'Brien}}{{Willingale} et~al.}{2013}]{wil13}
{Willingale} R.,  {Starling} R.~L.~C.,  {Beardmore} A.~P.,  {Tanvir} N.~R.,
  {O'Brien} P.~T.,  2013, \mn@doi [\mnras] {10.1093/mnras/stt175}, \href
  {http://adsabs.harvard.edu/abs/2013MNRAS.431..394W} {431, 394}

\bibitem[\protect\citeauthoryear{{Witthoeft}, {Bautista}, {Mendoza}, {Kallman},
  {Palmeri}  \& {Quinet}}{{Witthoeft} et~al.}{2009}]{wit09}
{Witthoeft} M.~C.,  {Bautista} M.~A.,  {Mendoza} C.,  {Kallman} T.~R.,
  {Palmeri} P.,   {Quinet} P.,  2009, \mn@doi [\apjs]
  {10.1088/0067-0049/182/1/127}, \href
  {http://adsabs.harvard.edu/abs/2009ApJS..182..127W} {182, 127}

\bibitem[\protect\citeauthoryear{{Wong} \& {Blitz}}{{Wong} \&
  {Blitz}}{2002}]{won02}
{Wong} T.,  {Blitz} L.,  2002, \mn@doi [\apj] {10.1086/339287}, \href
  {https://ui.adsabs.harvard.edu/abs/2002ApJ...569..157W} {569, 157}

\bibitem[\protect\citeauthoryear{{Yao}, {Schulz}, {Gu}, {Nowak}  \&
  {Canizares}}{{Yao} et~al.}{2009}]{yao09}
{Yao} Y.,  {Schulz} N.~S.,  {Gu} M.~F.,  {Nowak} M.~A.,   {Canizares} C.~R.,
  2009, \mn@doi [\apj] {10.1088/0004-637X/696/2/1418}, \href
  {http://adsabs.harvard.edu/abs/2009ApJ...696.1418Y} {696, 1418}

\bibitem[\protect\citeauthoryear{{Zeegers}, {Costantini}, {Rogantini}, {de
  Vries}, {Mutschke}, {Mohr}, {de Groot}  \& {Tielens}}{{Zeegers}
  et~al.}{2019}]{zee19}
{Zeegers} S.~T.,  {Costantini} E.,  {Rogantini} D.,  {de Vries} C.~P.,
  {Mutschke} H.,  {Mohr} P.,  {de Groot} F.,   {Tielens} A.~G.~G.~M.,  2019,
  \mn@doi [\aap] {10.1051/0004-6361/201935050}, \href
  {https://ui.adsabs.harvard.edu/abs/2019A&A...627A..16Z} {627, A16}

\bibitem[\protect\citeauthoryear{{Zhukovska}, {Dobbs}, {Jenkins}  \&
  {Klessen}}{{Zhukovska} et~al.}{2016}]{zhu16}
{Zhukovska} S.,  {Dobbs} C.,  {Jenkins} E.~B.,   {Klessen} R.~S.,  2016,
  \mn@doi [\apj] {10.3847/0004-637X/831/2/147}, \href
  {https://ui.adsabs.harvard.edu/abs/2016ApJ...831..147Z} {831, 147}

\makeatother
\end{thebibliography}

\end{document}